\documentclass[aps,prb,twocolumn,nofootinbib,superscriptaddress]{revtex4-2}
\usepackage{graphicx}
\usepackage{amsmath,amssymb}
\usepackage[utf8]{inputenc}
\usepackage{xcolor}
\usepackage{ulem}
\usepackage{bbm}
\usepackage[colorlinks = true,
linkcolor = blue,
urlcolor  = blue,
citecolor = blue,
anchorcolor = blue]{hyperref}
\usepackage{braket}
\usepackage{booktabs}
\usepackage{enumitem}
\usepackage[mathscr]{euscript}
\usepackage{mathdots}
\usepackage{color}
\newcommand{\sqt}{\sqrt{2}}

\begin{document}
\title{Floquet generation of hybrid-order topology and $\mathbb{Z}_2$-like bipolar localization}

\author{Koustav Roy$^@$}
\email{koustav.roy@iitg.ac.in}
\affiliation{Department of Physics, Indian Institute of Technology Guwahati-Guwahati, 781039 Assam, India}

\author{Latu Kalita$^@$}
\email{latukalita591@gmail.com}
\affiliation{Department of Physics, Indian Institute of Technology Guwahati-Guwahati, 781039 Assam, India}

\author{B. Tanatar}
\email[Corresponding author: ]{tanatar@fen.bilkent.edu.tr}
\affiliation{Department of Physics, Bilkent University, 06800 Bilkent, Ankara, T{\"u}rkiye}
\author{Saurabh Basu}
\affiliation{Department of Physics, Indian Institute of Technology Guwahati-Guwahati, 781039 Assam, India}

\begin{abstract}

Periodic driving offers a powerful tool to engineer topological phases and even induce phase transitions that are inaccessible in static settings. In this work, we demonstrate that a suitable driving protocol applied to the Benalcazar-Bernevig-Hughes (BBH) model, a canonical quadrupolar insulator with a $\pi$-flux-induced projective $\mathcal{PT}$-symmetry, gives rise to a \textit{hybrid-order} topological phase in which the dispersive first-order edge states and localized higher-order corner modes are shown to coexist at distinct quasienergies. Further, extending the scenario to non-reciprocal hopping induced non-Hermiticity, we uncover a $\mathbb{Z}_2$-like skin effect characterized by a sharp transition from a unipolar to a bipolar eigenstate‑localized phase. Remarkably, this phenomenon, which is established as a prerogative for spinful systems, emerges here purely from the interplay between the embedded gauge structure and Floquet-renormalized symmetry constraints, without invoking physical spin degrees of freedom. Further, the broken bulk-boundary correspondence in this driven non-Hermitian setting, can be restored via computing the two-dimensional generalized Brillouin zone (GBZ), which we construct through a symmetry-reduced mapping onto an effective one-dimensional problem. The resulting non-Bloch invariants faithfully capture both the higher-order topology and the unipolar to bipolar transition. These findings reveal periodic driving as a versatile and controllable handle for sculpting the interplay of higher-order topology, symmetry transmutation, and non-Hermitian skin physics in a single unified platform.
\end{abstract}

\maketitle
\def\thefootnote{@}\footnotetext{These authors contributed equally to this work}

%

\section{\label{s1}Introduction}
The Altland--Zirnbauer (AZ) classification \cite{Su1979,Bansil2016,Chiu2016}
provides a systematic foundation for characterizing topological phases of matter,
wherein nontrivial bulk invariants guarantee the existence of robust boundary states
through the bulk--boundary correspondence (BBC)
\cite{Hasan2010,Qi2011,Elliott2015,Armitage2018,Lv2021,Benalcazar2017Prime,Song2017,
Langbehn2017,Schindler2018}. The inclusion of crystalline symmetries further
expands this topological landscape, giving rise to higher-order topological phases
that host protected modes at the lower-dimensional boundaries, such as hinges and corners \cite{Schindler2018,Schindler2018Prime,Khalaf2018,Matsugatani2018,Franca2018,
Noguchi2021,Lahiri2024Prime,Lahiri2024DoublePrime,Ahn2019,Alase2016PRL,Ozawa2019,Yang2015,Wang2024}. A paradigmatic example is the Benalcazar-Bernevig-Hughes (BBH) model
\cite{Benalcazar2017,Lahiri2024,Wu2021,Liu2019}, a quantized
quadrupole insulator protected by mirror or inversion symmetry that,
despite belonging to symmetry class BDI (first-order trivial in two
dimensions), hosts robust zero-dimensional corner states in a square
geometry \cite{Schindler2018,Benalcazar2022}.
Beyond its role as a prototypical higher-order topological insulator,
the BBH model contains an embedded $\pi$-flux per plaquette
interpretable as an effective $\mathbb{Z}_2$ gauge field
\cite{Zhao2021,Li2022,Meng2023}, which modifies the antiunitary
symmetry algebra and produces an unconventional realization of
parity--time ($\mathcal{PT}$) symmetry
\cite{Zhao2016,Zhao2020,Zhao2021,Takahashi2024}.
In conventional systems, $(\mathcal{PT})^2= \pm 1$, where `$+$' sign applies to spinless particles and `$-$' for the spinful ones, the latter underpinning
Kramers degeneracy and helical edge modes.
Remarkably, the $\mathbb{Z}_2$ gauge structure reverses this
correspondence: a spinless BBH lattice realizes a spinful-like algebra
with $(\mathcal{PT})^2=-1$, while a spinful variant may satisfy
$(\mathcal{PT})^2=+1$, a symmetry transmutation driven purely by
lattice geometry without invoking physical spin. Nevertheless, the static BBH model remains first-order trivial in two
dimensions. The spinful-like $\mathcal{PT}$ algebra alone does not
generate dispersive edge modes or other hallmarks of spinful topological
phases. Thus, genuinely spinful boundary phenomena \cite{Okuma2020,Liu2020} remain absent in
the static regime.
This tension between an unconventional symmetry algebra and a
constrained bulk topological response raises a fundamental question:
can one dynamically transfigure a higher-order crystalline insulator
into a phase that simultaneously hosts conventional (first) and
higher-order boundary modes, without altering its microscopic
symmetries?

Periodically driven systems offer a precise route to overcome
such a limitation \cite{Grifoni1998,Restrepo2016,Goldman2014}.
The quasienergy spectrum, permits
independent topological gaps at both zero and $\pi/T$ quasienergies, so
that anomalous boundary modes can emerge even when every instantaneous
Hamiltonian within the driving cycle is trivial
\cite{Cayssol2013,Rudner2013,GomezLeon2013}.
Crucially, temporal modulation can also reorganize the effective
symmetry constraints of the Floquet operator, facilitating modified boundary
phenomena strictly forbidden in the static limit
\cite{Roy2023,Jangjan2022,Yang2022,Roy2024Prime,Roy2024DoublePrime}.
Motivated by these prospects, we investigate a periodically driven BBH
lattice and demonstrate that appropriate driving protocols dynamically
activate first-order topology while preserving the higher-order
quadrupole character, yielding a \textit{hybrid-order} topological
phase in which the dispersive edge states coexist with
symmetry-protected corner modes in distinct quasienergy sectors,
an arrangement fundamentally inaccessible in the static limit.

The landscape becomes even richer once Hermiticity is relinquished, since
realistic platforms inevitably involve gain, loss, or non-reciprocal
transport
\cite{Lee2016,Shen2018,Ghatak2019,Gong2018,Kawabata2019,Ashida2020,Bergholtz2021,Banerjee2023,Borgnia2020}.
In such systems the non-Hermitian skin effect (NHSE)  wherein a
macroscopic number of bulk eigenstates accumulate exponentially at
system boundaries
\cite{Yao2018,Kunst2018,Lee2019,Borgnia2020,Okuma2020Prime,Hetenyi2025PRB}  can
invalidate the conventional BBC, and in spinful systems symmetry
constraints further stabilize $\mathbb{Z}_2$ skin effects with
bipolar localization \cite{Okuma2020,Okuma2023,Zhang2022,Lin2023}.
Extending our driven BBH framework to a non-reciprocal setting, we
uncover a striking $\mathbb{Z}_2$-like skin effect within an
effectively spinless platform.
Periodic driving further controls a transition between unipolar
localization (all eigenstates accumulating at a single corner) to
bipolar localization (symmetry-related sectors at opposite corners)
\cite{Roy2025}, and can even entirely suppress the skin effect,
restoring an extended bulk spectrum.
Since the pronounced spectral migration invalidates the standard Bloch
framework, a consistent topological characterization requires the
non-Bloch formalism with topology defined on a generalized Brillouin
zone (GBZ) \cite{Yokomizo2019,Zhang2020,Wu2020}.
Although GBZ construction is well established in one dimension
\cite{Yokomizo2021,Roy2025Prime,Roy2025DoublePrime,Zhang2020Prime,Wu2020},
extending it to two dimensions is significantly more challenging. Here, we
exploit the underlying mirror symmetry to reduce the 2D problem to an
effective 1D representation, enabling construction of the Floquet GBZ
and the corresponding NH topological invariants.
\par Put together, our study establishes periodic driving as a unifying framework that (i) activates first-order topology in a higher-order trivial lattice, (ii) enables coexistence of edge and corner states, (iii) realizes spinful symmetry algebra within an effectively spinless platform, and (iv) controls NH skin phenomena, including $\mathbb{Z}_2$-like localization and unipolar–bipolar transitions. More broadly, our results reveal how the interplay of crystalline symmetry, Floquet engineering, and NH reshapes symmetry classification and boundary correspondence beyond equilibrium scenarios.

\par The remainder of this paper is organized as follows: Sec.~\ref{s2} introduces the model and the emergent $\mathcal{PT}$
symmetry algebra; Sec.~\ref{s3} presents the driving protocol and emergence of \textit{hybrid-order}
phase;  Sec.~\ref{s4} analyzes the NH unipolar--bipolar skin transition, and develops the non-Bloch Floquet GBZ and topological
invariants; Sec. \ref{s5_expt} proposes a topolectrical platform for the driven model, and Sec.~\ref{s6} summarizes the main findings and provides concluding remarks. Finally, Appendices~\ref{app1} and \ref{app2} provide, respectively, a detailed 
account of the drive-induced symmetry transmutation and the derivation 
of the effective $\mathbf{d}$-vectors used in the Floquet GBZ construction.

\begin{figure}[t]
         \includegraphics[width=0.95\columnwidth]{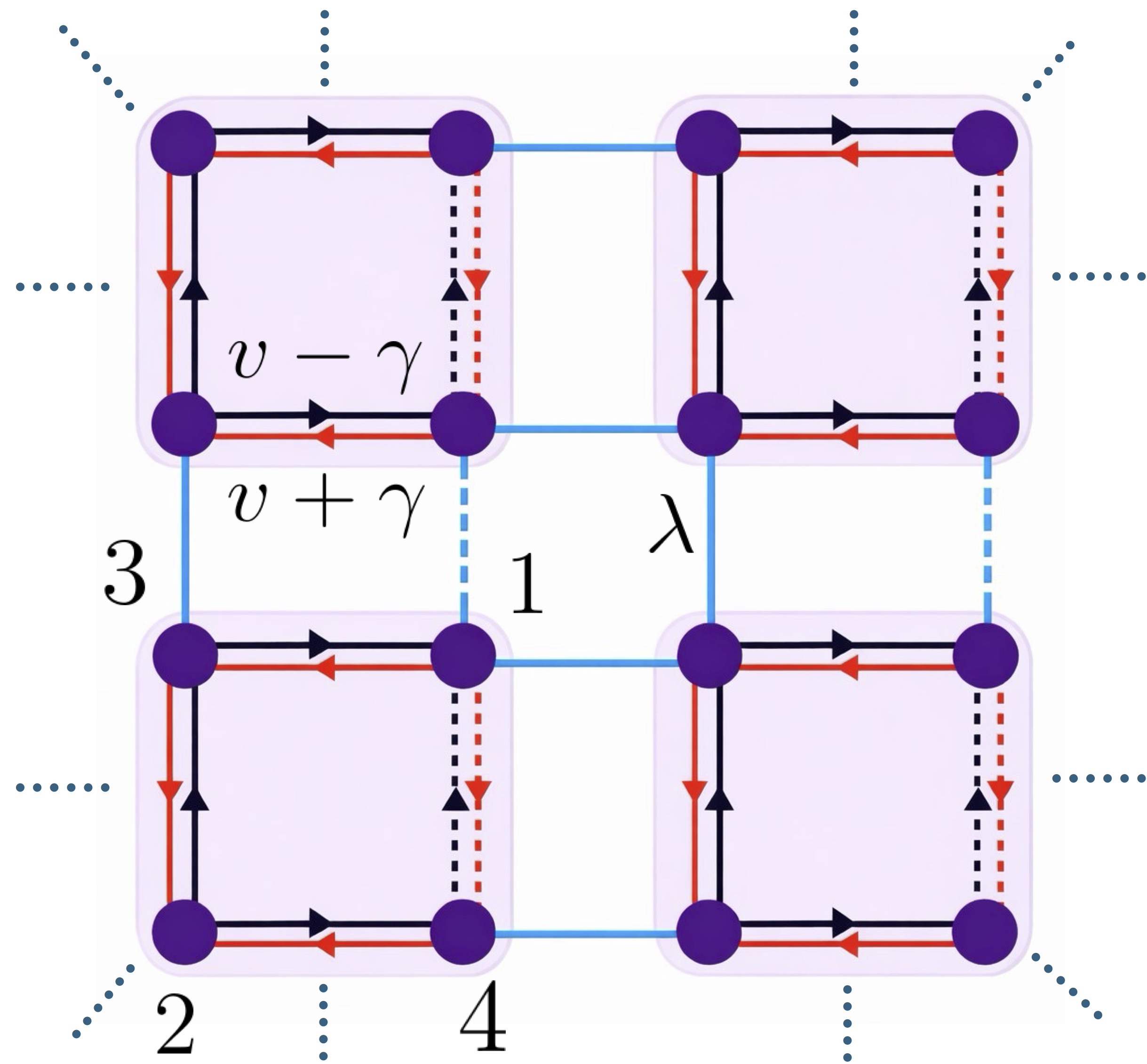}
\caption{Schematic illustration of an $N \times N$ square lattice with intercell hopping amplitude $\lambda$ and nonreciprocal intracell hopping amplitudes $v \pm \gamma$. The dashed bonds indicate hopping processes that differ by a $\pi$ phase relative to the solid bonds. This staggered phase pattern effectively introduces a $\mathbb{Z}_2$ gauge structure, which modifies the symmetry algebra and consequently alters the topological classification of the system.
}
\label{1}
\end{figure}
\section{\label{s2}Model Hamiltonian and Emergent $\mathcal{PT}$ Symmetry}
We consider a minimal tight-binding realization of a two-dimensional second-order topological insulator on a square lattice (as shown in Fig. \ref{1}), where each unit cell accommodates four internal (orbital) degrees of freedom. The lattice geometry incorporates asymmetric intracell hoppings together with a staggered phase configuration that effectively inserts a $\pi$ flux through every plaquette (as seen in Fig. \ref{1}, where the dashed lines denote the hopping amplitudes with a $\pi$-phase difference
from their solid counterparts). This flux pattern plays a central role, endowing the model with a nontrivial $\mathbb{Z}_2$ gauge structure that subtly reshapes its symmetry properties. The corresponding real-space Hamiltonian takes the form \cite{Benalcazar2017,Lahiri2024,Wu2021,Liu2019},
\begin{align}
H_{2D} =  & \sum_{\mathbf{n}} \Big\{ (v - \gamma)\left[
c^{\dagger}_{\mathbf{n},1}(c_{\mathbf{n},3} - c_{\mathbf{n},4})
+ (c^{\dagger}_{\mathbf{n},3} + c^{\dagger}_{\mathbf{n},4}) c_{\mathbf{n},2}
\right] \nonumber \\
&+ (v + \gamma)\left[
(c^{\dagger}_{\mathbf{n},3} - c^{\dagger}_{\mathbf{n},4}) c_{\mathbf{n},1}
+ c^{\dagger}_{\mathbf{n},2}(c_{\mathbf{n},4} + c_{\mathbf{n},3})
\right] \nonumber \\
&+ \lambda \Big[
c^{\dagger}_{\mathbf{n},1}(c_{\mathbf{n}+\hat{x},3} - c_{\mathbf{n}+\hat{y},4})
\nonumber \\
& 
+ c^{\dagger}_{\mathbf{n},2}(c_{\mathbf{n}-\hat{y},3} + c_{\mathbf{n}-\hat{x},4})
+ \mathrm{H.c.}
\Big]
\Big\},
\end{align}
 where $c^{\dagger}_{\mathbf{n},j}$ creates a fermion on sublattice 
$j = 1,2,3,4$ within unit cell 
$\mathbf{n} = (n_x, n_y)$, and $\hat{x}, \hat{y}$ denote the primitive lattice vectors. 
The parameter $\lambda$ controls the intercell hopping, while 
$v \pm \gamma$ describe the intracell amplitudes, with 
$\gamma$ encoding the non-reciprocity. In momentum space, defining 
$\psi_{\mathbf{k}} = (c_{\mathbf{k},1}, c_{\mathbf{k},2}, 
c_{\mathbf{k},3}, c_{\mathbf{k},4})^{\mathsf{T}}$, 
the Bloch Hamiltonian assumes a compact form,
\begin{equation}
\begin{split}
H_{2D}(\mathbf{k})  = & \; [v + \lambda \cos(k_x)] \tau_x \sigma_0 
- [\lambda \sin(k_x) + i\gamma] \tau_y \sigma_z \\
&+ [v + \lambda \cos(k_y)] \tau_y \sigma_y 
+ [\lambda \sin(k_y) + i\gamma] \tau_y \sigma_x .
\end{split}
\end{equation}
where $\tau_{\mu}$ and $\sigma_{\mu}$ are the Pauli matrices acting on orbital subspaces.

\par At this stage, in order to clarify its intrinsic topological character, we first focus on the Hermitian limit, $\gamma=0$. In this regime, the model realizes a quadrupolar insulator supporting zero-energy corner states for $|v/\lambda| <1$, and characterized by a quantized bulk quadrupole moment. Importantly, this higher-order topology is protected by crystalline symmetry rather than internal symmetries alone. In particular, the Hamiltonian respects the mirror-rotation symmetry \cite{Liu2019},
\begin{equation}
M_{xy}\, H(k_x,k_y)\, M_{xy}^{-1} = H(k_y,k_x),
\end{equation}
with
\begin{equation}
M_{xy} = \frac{(\tau_0 - \tau_z)\sigma_x - (\tau_0 + \tau_z)\sigma_z}{2}.
\label{eq:mirror_symmetry}
\end{equation}

It further obeys chiral symmetry
\begin{equation}
S\, H(\mathbf{k})\, S^{-1} = -H(\mathbf{k}), 
\qquad
S = \tau_z \sigma_0,
\label{eq:chiral_sym}
\end{equation}
alongside intrinsic particle-hole symmetry 
$C = \tau_z \sigma_0 K$ 
and time-reversal symmetry 
$\mathcal{T} = K$($K$ denoting complex conjugation). Accordingly, the model belongs to class BDI in the AZ classification \cite{Chiu2016}. In two spatial dimensions, this symmetry class is first-order trivial, explaining the absence of symmetry-protected dispersive edge states despite the protection by crystalline symmetries. 
\par Nevertheless, a complete symmetry characterization requires extending the framework 
to include inversion symmetry and the associated AZ$+I$ classification 
based on $\mathcal{PC}$, and $\mathcal{PT}$ \cite{Takahashi2024}. Interestingly, in our system, the symmetry analysis acquires an unexpected twist once inversion symmetry is considered in the presence of the embedded $\pi$-flux. In conventional systems, the combined space-time inversion satisfies 
$(\mathcal{PT})^2 = +1$ for spinless particles and 
$(\mathcal{PT})^2 = -1$ for spin-1/2 systems, 
following the general relation 
$(\mathcal{PT})^2 = (-1)^{2s}$.
However, the $\pi$-flux configuration renders the inversion symmetry inherently projective. To remain compatible with the gauge structure, inversion, therefore, must be redefined as \cite{Zhao2021,Zhao2016},
\begin{equation}
    \tilde{P} = G\mathcal{P}
\end{equation}
where $G$ is a gauge transformation satisfying 
\begin{equation}
[G,\mathcal{T}] = 0, \qquad \{G,\mathcal{P}\} = 0, \qquad G^2 = 1.
\label{eq:gauge_condition}
\end{equation}
As a result,
$\tilde{P}^{\,2}=(G\mathcal{P})^2=-1$,
and the combined space-inversion symmetry gets generalized as,
$(\tilde{P}\mathcal{T})^2 = (-1)^{2s+1}$.
In our model, the appropriate gauge transformation is $\tau_0 \sigma_z$, which commutes with time-reversal while anticommuting with inversion. Thus, give rise to a redefined $\mathcal{PT}$ algebra, represented as, $\tilde{P} \mathcal{T} = G \mathcal{PT} = i \tau_0 \sigma_y K$, (with the standard choice of inversion symmetry being $\mathcal{P}=\tau_0 \sigma_x$) which obeys $(\tilde{P}\mathcal{T})^2=-1$. Consequently, even though the microscopic degrees of freedom are spinless, the system satisfies a spinful-like algebra with $(\tilde{P}\mathcal{T})^2=-1$. The $\pi$-flux therefore induces a symmetry transmutation, whereby the distinction between spinless and spinful algebra is effectively inverted. In this sense, the lattice realizes a synthetic spinful symmetry structure without introducing `real' spin.
\par Nonetheless, this algebraic restructuring does not automatically generate first-order topology. Even within the extended AZ+$I$ classification, despite exhibiting a spinful-like $\mathcal{PT}$ structure, the static model remains first-order trivial in two dimensions. The quadrupole phase survives as a crystalline higher-order state, while symmetry-protected dispersive edge modes remain absent. Thus, although the $\pi$-flux profoundly alters the symmetry algebra, the topological classification itself remains unaltered in the static setting. This observation naturally motivates the question of whether one can relax the effective $(\mathcal{PT})^2$ constraint, without breaking the symmetry itself, and thereby unlock first-order topology in a system that is otherwise restricted to higher-order behavior. As we demonstrate in the following section, periodic driving provides precisely such a mechanism, enabling a re-organization of symmetry constraints and activating topological phenomena inaccessible in the static limit.
\begin{figure*}[t]
\centering
\includegraphics[width=0.92\textwidth]{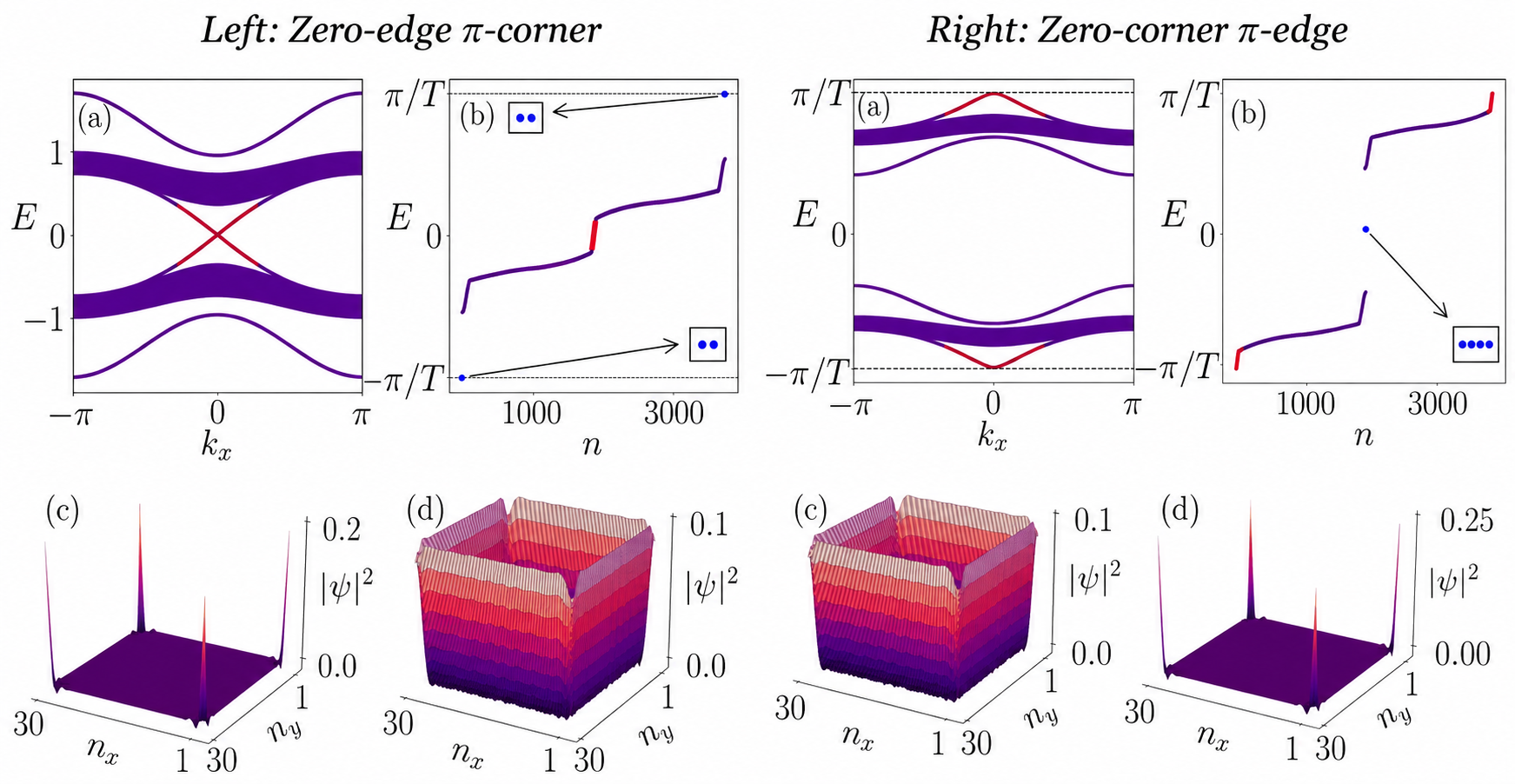}
\caption{Floquet hybrid-order topological phase with coexisting edge and corner states. Left panel: Quasienergy spectra under open boundary conditions along $y$ (a) and along both $x$ and $y$ (b), for $T_1=1$, $T_2=4$, $\lambda=0.5$, $v=2.5$, $N_x=N_y=30$, $\gamma=0$. Red branches denote gapless first-order boundary states; blue dots denote localized second-order corner states. Edge modes emerge within the quasienergy gap at zero, while the corner states remain pinned at $\pi/T$. Panels (c) and (d) show the spatial probability distributions for the $\pi/T$ corner states and zero-quasienergy edge states, respectively, confirming the coexistence of conducting edges and localized corners --- the hybrid-order phase. Right panel: Same quantities as the left panel, but at $v=0.5$ (all other parameters unchanged). The localization pattern is reversed relative to the figure in the left panel: edge modes now emerge at $\pi/T$ [panel (a), red] while corner states pin to zero quasienergy [panel (b)]. Panels (c) and (d) confirm the corresponding spatial character.}
\label{2}
\end{figure*}
\section{\label{s3}Driven model and its symmetries}
To dynamically reshape the symmetry-constrained topology of the static BBH model, we implement a Floquet engineering protocol designed to interconvert $\mathcal{PT}$-symmetric topological phases in the presence of the embedded $\mathbb{Z}_2$ gauge structure. A central requirement of the driving protocol is that periodic modulation preserves the fundamental $\mathcal{PT}$ symmetry of the system while allowing its algebraic realization and consequently the associated topological classification, to be modified dynamically. To this end, we consider a step-driving scheme in which intracell and intercell hopping processes are activated sequentially within a single driving period. The time-dependent Hamiltonian is written as
\begin{equation}
\mathcal{H}(\mathbf{k}, t) =
\begin{cases}
\mathcal{H}_1(\mathbf{k}), & t \in [nT,\, nT+T_1), \\
\mathcal{H}_2(\mathbf{k}), & t \in [nT+T_1,\,(n+1)T),
\end{cases}
\end{equation}
where $\mathcal{H}_1 = H(v,\lambda=0)$ describes purely intracell hopping and 
$\mathcal{H}_2 = H(v=0,\lambda)$ governs intercell tunneling, with 
$n\in\mathbb{Z}$ and $T=T_1+T_2$ denoting the total driving period. Within Floquet theory, the stroboscopic dynamics are captured by the one-period evolution operator
\begin{equation}
U(T)=
e^{-i\mathcal{H}_2 T_2}
e^{-i\mathcal{H}_1 T_1},
\end{equation}
which defines the effective Floquet Hamiltonian, $\mathcal{H}_{\text{eff}}=\frac{i}{T}\ln U(T).$ The eigenvalues of $\mathcal{H}_{\text{eff}}$, defined modulo $2\pi/T$, constitute the quasienergy spectrum governing the nonequilibrium topology of the driven system.
\par However, unlike the static Hamiltonian, periodically driven systems generally do not preserve static symmetry operations in a straightforward manner, since the Hamiltonians at different times fails to commute. Consequently, periodic driving may obscure the internal symmetries present in the static model. Thus, in order to restore the symmetry structure, we introduce \emph{symmetric time frames} \cite{Asboth2013,Asboth2014,Wang2021} generated by the unitary transformations,
\begin{equation}
F_1=e^{-i\mathcal{H}_1T_1/2},
\qquad
F_2=e^{\,i\mathcal{H}_2T_2/2},
\label{eq:symmetric_frame}
\end{equation}
which define equivalent Floquet operators
$U_j = F_jU(T)F_j^{-1}$ ($j=1,2)$. While these operators retain the same quasienergy spectrum similar to $\mathcal{\mathcal{H_{\text{eff}}}}$, they permit a consistent redefinition of symmetry transformations within the driven setting. In particular, time-reversal and chiral symmetries re-emerge through Floquet-renormalized operators,
\begin{equation}
\mathcal{T}_F = F_j^{-1}(-\mathbf{k})\,\mathcal{T}\,F_j(\mathbf{k}),
\qquad
S_F = F_j^{-1}(\mathbf{k})\,S\,F_j(\mathbf{k}),
\end{equation}
such that the effective Hamiltonian satisfies
\begin{align}
\mathcal{T}_F \mathcal{H}_{\text{eff}}(\mathbf{k}) \mathcal{T}_F^{-1}
&= \mathcal{H}_{\text{eff}}^{\dagger}(-\mathbf{k}), \\
S_F \mathcal{H}_{\text{eff}}(\mathbf{k}) S_F^{-1}
&= -\mathcal{H}_{\text{eff}}(\mathbf{k}).
\end{align}
Thus, the periodically driven model retains the essential symmetry content of its static counterpart, albeit in a dynamically renormalized form. A more detailed discussion of the drive-induced symmetry reformulation is provided in Appendix \ref{app1}.
\par A decisive consequence of this dynamical reconstruction is the breakdown of the gauge-induced constraint associated with the embedded $\mathbb{Z}_2$ structure. It turns out that the Floquet-renormalized symmetry operators no longer respect the commutation relations (as shown in Eq. \ref{eq:gauge_condition}) required by the underlying $\mathbb{Z}_2$ gauge structure. In particular, one finds $[G,\mathcal{T}_F]\neq0$, implying that the projective inversion symmetry $\tilde{P}=G\mathcal{P}$ no longer remains as a valid symmetry operation in the driven setting. Consequently, periodic driving therefore dynamically unleashes the system from the projective symmetry algebra via dynamically breaking the $\mathbb{Z}_2$ gauge constraint thus restoring the conventional relation, $(\mathcal{PT})^2=+1$, which results in $\mathcal{P}=I$, with $I$ being the momentum inversion operator. In this manner, the driven lattice realizes an interconversion between distinct $\mathcal{PT}$-symmetric topological sectors through dynamical modification of the gauge structure rather than explicit symmetry breaking. In this way, although the standard AZ classification remains formally unchanged (suggesting first-order triviality) the modified AZ+$I$ framework based on $\mathcal{PT}$ and $\mathcal{PC}$ symmetries predicts a fundamentally different outcome. Under the restored 
$\mathcal{PT}$ formalism, the Floquet Hamiltonian enters a symmetry class capable of supporting first-order dispersive edge modes without harming the preexisting higher-order topology \cite{Takahashi2024}. Consequently, the periodically driven BBH model hosts a hybrid-order topological phase, characterized by the simultaneous presence of dispersive edge states and symmetry-protected corner modes.
\subsection{\textbf{Drive-Induced Hybrid Order Topology}}
\par We substantiate this prediction through numerical analysis of the quasienergy spectrum under both periodic and open boundary conditions. For parameters $T_1=1$, $T_2=4$, $\lambda=0.5$, $N_x=N_y=30$, and $\gamma=0$, the cylindrical geometry as shown in the left panel of Fig. \ref{2}(a), reveals gapless boundary modes around zero quasienergy for an arbitrary value of $v$, say $v=2.5$, while a finite gap persists near the $\pi/T$ quasienergy sector. Upon imposing open boundaries along both spatial directions (left panel of Fig. \ref{2}(b)), four symmetry-protected states emerge precisely at quasienergy $\pi/T$, localized at the four corners of the finite sample. Their spatial character is confirmed through probability-density distributions on a $30\times 30$ lattice, as shown in the left panel of Figs. \ref{2}(c)-(d), demonstrating corner confinement, whereas the zero-quasienergy modes remain extended along system edges. Remarkably, the two distinct topological orders occupy distinct quasienergy sectors, for example, edge modes reside at zero quasienergy, while corner modes appear at $\pi/T$, highlighting the intrinsically out-of-equilibrium origin of the hybrid phase. A complementary regime further illustrates this dynamical flexibility. For instance, at $v=0.5$ (with all other parameters staying unchanged) the localization pattern reverses, with the corner states migrating to the zero-quasienergy sector while the dispersive edge modes appearing near the $\pi/T$ quasienergy (as shown in the right panel of Fig. \ref{2}). Periodic driving therefore enables controlled redistribution of topological order between quasienergy gaps, establishing a dynamically generated hybrid phase in which first- and second-order topology coexist, an arrangement fundamentally inaccessible within static $\mathcal{PT}$-symmetric lattices.
\subsection{\label{s3b}\textbf{Characterization of the Hybrid Order Topology}}
\subsubsection{\textbf{Why no momentum-space invariant works}}

Having established the coexistence of first- and second-order boundary modes in the hybrid-order Floquet phase, we now focus on characterizing these states more rigorously. Since the edge and corner modes originate from a genuine symmetry transmutation of the driven lattice rather than from an accidental band-touching, it is essential to confirm their robustness by identifying a topological invariant that protects them. It turns out that the obvious candidates fail. For example, the ordinary Chern number vanishes, since the time-reversal symmetry in the Floquet-renormalized frames makes the Berry curvature odd; furthermore, the Kane--Mele $\mathbb{Z}_2$ invariant is not relevant as well, as it requires a spinful Kramers structure with $\mathcal{T}^2=-1$, whereas our internal degrees of freedom are orbital, with $\mathcal{T}=K$ and $\mathcal{T}^2=+1$.

The relevant symmetry is once again the space--time inversion $\mathcal{PT}$, whose algebra dictates the appropriate invariant. As described earlier, in the conventional frame, the drive restores $(\mathcal{PT})^2=+1$, indicating a real Chern number, obtained from the Stiefel--Whitney classes of the resulting real bundle \cite{Zhao2016,Zhao2021,Zhao2020}. This invariant however fails, since the first- and second-order phases never share same quasienergy gap. For instance, when the edge states occupy the zero gap the corner modes pin to $\pi/T$, and tuning the parameter $v$ interchanges them (Fig. \ref{2}). Therefore, the topological protection is a property of each gap \emph{separately}. Any band invariant, the real Chern number included, measures only the net edge-state difference across a single gap and cannot resolve either gap individually when both are topologically protected, the same mechanism by which the ordinary Chern number vanishes in anomalous Floquet phases with robust boundary modes in each gap \cite{Rudner2013}.

This failure motivates a shift to the symmetric time frames of Appendix~\ref{app1} (as well as discussed in Sec \ref{s3}), where the composite antiunitary operator $\mathcal{A} = i\,\tau_0\sigma_y K$ commutes with both step Hamiltonians and satisfies $\mathcal{A}\,U_{1,2}\,\mathcal{A}^{-1} = U_{1,2}^{-1}$, ($j=1,2$ represents the two symmetric frames) but now squares to $\mathcal{A}^2=-1$. By Kramers' theorem this forces an exact two-fold degeneracy of the quasienergy spectrum at every $\mathbf{k}$, $\{-E,-E,+E,+E\}$, and closes off the momentum-space route for enumerating the topological invariant entirely, since the Stiefel--Whitney classes, the Euler class, and the real Chern number all require a momentum-local antiunitary squaring to $+1$. Whereas $\mathcal{A}^2=-1$ renders the bundle quaternionic and these invariants strictly undefined. We therefore turn to a real-space construction, namely the spectral localizer, capable of resolving each gap independently, developed below.
\subsubsection{\textbf{Real-Space Characterization: The Spectral Localizer and the Chiral Marker}}

A real-space marker extracts a quantized topological number directly from the Hamiltonian of a finite, open sample together with its position operators, with no reference to momentum space or translational invariance. The tool we employ for this purpose is the spectral localizer, built from the position operators $X,Y$ and the Hamiltonian as \cite{Cerjan2022,Lahiri2025}
\begin{equation}
  \mathcal{L}(x_0,y_0,E) = \kappa\,\sigma_x\otimes(X-x_0) + \kappa\,\sigma_y\otimes(Y-y_0) + \sigma_z\otimes(H-E),
  \label{eq:loc}
\end{equation}
where $(X-x_0,Y-y_0)$ measures displacement from a probe point $(x_0,y_0)$, while $(H-E)$ measures displacement from a probe energy. The Pauli matrices combine these non-commuting operators into a single Hermitian matrix on an auxiliary two-level space, and $\kappa>0$ balances the relative weight of position and energy. At this point two useful quantities follow from $\mathcal{L}$. First, the \emph{localizer gap} $\sigma^{\mathcal{L}}_{\min}(x_0,y_0,E)$, which is the smallest singular value of $\mathcal{L}$, and vanishes precisely where a protected state resides at that position and energy, so that scanning the probe point produces a spatial map of the protected states; and second the \emph{signature} $\mathrm{Sig}\,\mathcal{L}$, which is the difference between the number of positive and negative eigenvalues of $\mathcal{L}$, and is quantized because it can change only when an eigenvalue crosses zero, i.e., only where the localizer gap closes.
\begin{figure}[t]
\centering
\includegraphics[width=1\columnwidth]{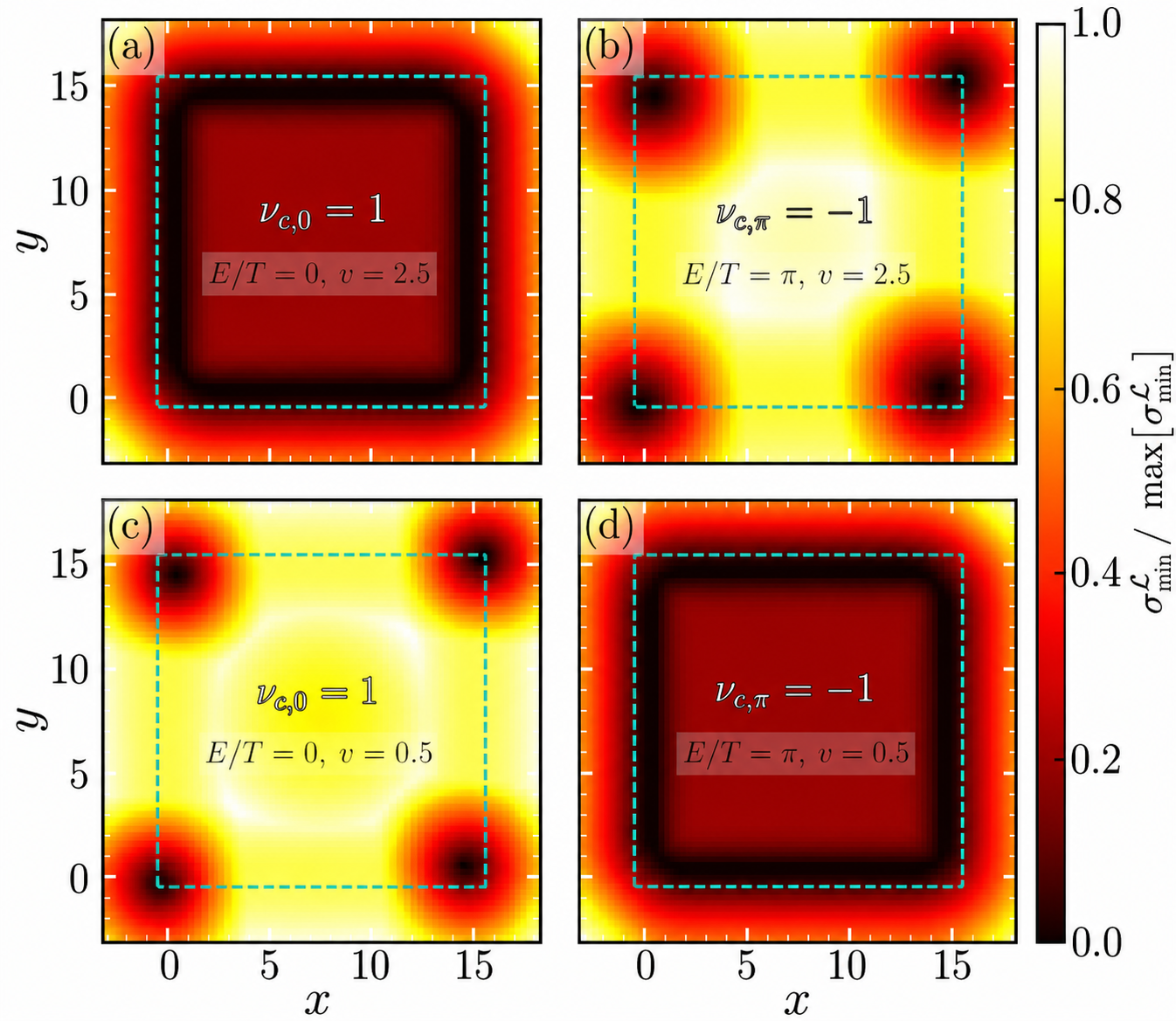}
\caption{Panels (a) and (d) display the spectral localizer gap closing along a continuous one-dimensional ring tracing the sample boundary, signaling dispersive first-order edge states, while panels (b) and (c) show closures only at the four sample corners, with edges and bulk remaining gapped, signaling second-order corner states. The localizer is evaluated across the sample plane at $E=0$ and $E=\pi/T$ for the parameter regimes of the left and right panels of Fig. \ref{2} respectively.}
\label{localizer vs disorder}
\end{figure}

Fig.~\ref{localizer vs disorder} shows $\sigma^{\mathcal{L}}_{\min}$ scanned across the sample plane at $E=0$ and $E=\pi/T$ for the two parameter regimes in the left and right panels of Fig.~\ref{2}, respectively. The resulting maps fall into two distinct classes: panels (a) and (d) show the localizer gap closing along a continuous one-dimensional ring tracing the full perimeter of the sample, the signature of dispersive first-order edge states, while panels (b) and (c) show closure only at four isolated points at the corners, with the edges and interior remaining gapped, the signature of localized second-order corner modes. This reproduces, entirely in real space, the central claim of the manuscript: the two quasienergy gaps carry different topological orders, and tuning $v$ interchanges them.

Hence the natural step is to compute the half-signature $\tfrac12\mathrm{Sig}\,\mathcal{L}$, i.e., the local Chern marker; however, this quantity is forced to vanish identically in our system, since the chiral symmetry ($S$ of Eq.~\eqref{eq:chiral_sym}) is on-site and therefore commutes with the position operators, $[S,X]=[S,Y]=0$, pairing every eigenvalue of $\mathcal{L}$ with its negative. A marker that exploits the chiral symmetry constructively, rather than being defeated by it, is therefore required, and this is furnished by the chiral marker. For our BBH lattice, with grading $S$ satisfying $[S,X]=[S,Y]=0$ in the symmetric frames, the appropriate construction is built along the lattice diagonal $\hat n = (\cos 45^\circ,\sin 45^\circ)$, i.e., along $X+Y$ --- the direction along which the sample corners lie and the symmetry axis of the model's mirror --- and is defined as
\begin{equation}
  \mathcal{A}_{\hat n} = \big[\kappa\big(\cos\theta\,(X-x_0)+\sin\theta\,(Y-y_0)\big) + i H_{\rm eff}\big]\,S,
  \label{eq:chiralmarker}
\end{equation}
with the corresponding marker $\nu_{\hat n} = \tfrac12\,\mathrm{Sig}\,\mathcal{A}_{\hat n}$, evaluated at $\theta = 45^\circ$, with the physically meaningful value referenced against vacuum.
\begin{figure}[t]
\centering
\includegraphics[width=1\columnwidth]{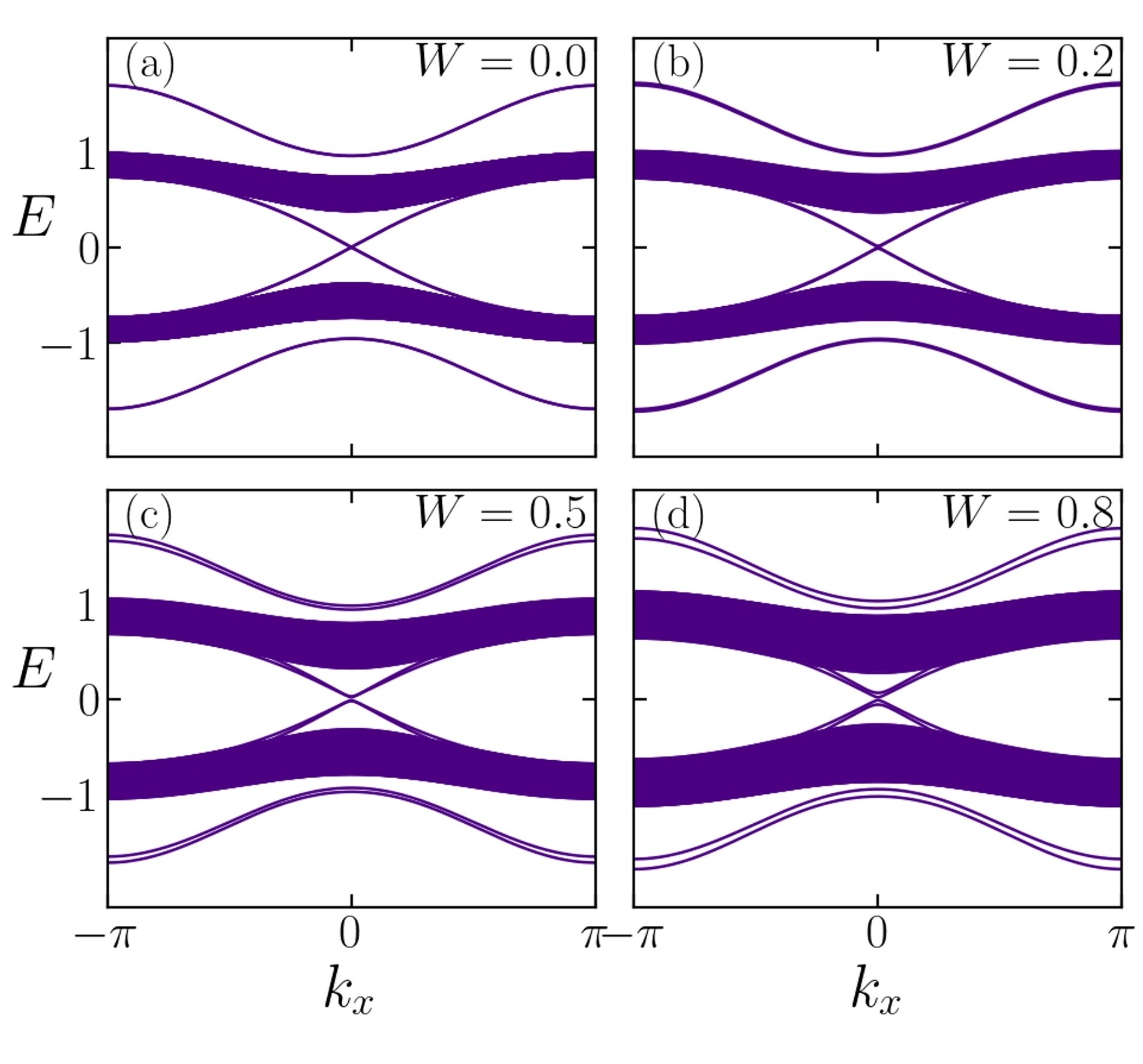}
\caption{Quasienergy spectrum under open boundary conditions along the $y$ direction and periodic boundary conditions along the $x$ direction, demonstrating the presence of drive-induced first-order edge states. Panels (a)--(d) show the evolution of the spectrum in the presence of random on-site disorder with disorder strengths $W=0$, $0.2$, $0.5$, and $0.8$, respectively. The edge states remain robust up to the critical disorder strength $W=0.5$, reflecting the persistence of the topological protection. For stronger disorder ($W=0.8$), the disorder dominates over the topological energy scales, lifting the edge-state degeneracy and destroying the protected edge modes.}
\label{disorder}
\end{figure}
Since the marker of Eq.~\eqref{eq:chiralmarker} requires the effective Hamiltonian to genuinely anticommute with $S$, and the conventional Floquet operator does not do so, the construction must be carried out in the symmetric frames of Eqs.~(A2)--(A3), the same frames in which the chiral symmetry is revived and in which $\mathcal{A}^2=-1$ (see the corresponding discussion in Sec. \ref{s3} and \ref{s3b} respectively. Frames 1 and 2 share the same quasienergies but are related by a unitary that does not commute with $S$. So the two frames carry independent topological information --- the two-dimensional analogue of a known result for one-dimensional chiral Floquet systems \cite{Asboth2014}. While each invariant alone cannot distinguish the different localized states, the Floquet classification \cite{Roy2017} allows the construction of a pair of physically meaningful invariants \cite{Asboth2013,Asboth2014},
\begin{equation}
  \nu_{c,0} = \tfrac12\big(\nu^1_{c,0}+\nu^2_{c,0}\big),
  \qquad
  \nu_{c,\pi} = \tfrac12\big(\nu^1_{c,\pi}-\nu^2_{c,\pi}\big),
  \label{eq:pair}
\end{equation}
where the subscript and the superscript denote the branch and the frame, respectively. At the representative point considered here we measure $(\nu^1_{c,0},\nu^1_{c,\pi}) = (+1,-1)$ in frame~1 and $(\nu^2_{c,0},\nu^2_{c,\pi}) = (+1,+1)$ in frame~2, yielding $(\nu_{c,0},\nu_{c,\pi}) = (+1,-1)$. The symmetric frames, originally forced upon us by the $(\mathcal{PT})^2=-1$ ($\mathcal{A}^2=-1$) algebra, thus supply exactly the two independent real-space invariants needed to characterize the two quasienergy gaps separately --- a resolution no single band invariant of the conventional frame could provide.
\par To further substantiate the robustness of the drive-induced first-order edge states, we examine the effect of random disorder, introduced through the intracell hopping amplitude $v \to v + \delta v_i$, with $\delta v_i$ drawn independently and uniformly from the interval $[-W/2,\,W/2]$ at each site $i$.
Fig.~\ref{disorder} shows the resulting quasienergy spectrum under open boundary conditions along $y$ and periodic boundary conditions along $x$, for the representative parameter point of panel~(a) of the left panel of Fig.~\ref{2} --- namely, the regime hosting zero-quasienergy edge states and $\pi$-quasienergy corner states; an entirely analogous disorder analysis can be carried out for the complementary configuration with the roles of the two gaps interchanged. As the disorder strength is increased, the edge states remain robust up to a critical value $W=0.5$, comparable to the intercellular hopping amplitude $\lambda=0.5$, a value used throughout the manuscript. Beyond this threshold, the disorder overwhelms the topological protection: the previously degenerate edge states split into non-degenerate modes, signalling the breakdown of the $\mathcal{PT}$ symmetry responsible for their protection, and the protected edge modes cease to survive at larger disorder strengths.
\par At this stage it is also important to emphasize that, throughout the above analysis, we have 
restricted ourselves to the purely Hermitian limit ($\gamma = 0$). This 
assumption was necessary, since NH topology cannot, in general, 
be fully characterized within the conventional AZ or AZ$+I$ 
symmetry classifications. In particular, the presence of non-reciprocity 
necessitates a non-Bloch framework for a consistent topological description. 
Nevertheless, the periodically driven system remains $\mathcal{PT}$ symmetric, 
and the emergence of dispersive edge modes continues to exhibit intriguing 
spinful characteristics even in the NH regime. In the following 
section, we shall demonstrate how periodic driving gives rise to distinctive 
NH phenomena specific to spinful systems, which can be 
systematically explored even within our spinless model.
\section{\label{s4}Non-Hermitian case: Emergent Higher order $\mathbb{Z}_2$ skin effect}
\subsection{Drive-Induced Unipolar-Bipolar Transition}
We now extend our analysis to the NH regime in order to investigate the fate of boundary localization once non-reciprocity is introduced into the periodically driven BBH lattice. Non-Hermiticity is incorporated through asymmetric intracell hopping amplitudes characterized by the parameter $\gamma$ (Fig.~\ref{1}), which induces directional particle transport and consequently gives rise to non-Hermitian skin accumulation. In the absence of periodic driving, the coexistence of inversion and mirror symmetries together with the underlying lattice geometry promotes a higher-order manifestation of the skin effect, wherein the bulk eigenstates accumulate at a single corner rather than along the one-dimensional edges. We refer to this phenomenon as a \textit{second-order skin effect} \cite{Lin2023,Okugawa2020}, reflecting its higher-codimension localization behavior.
Remarkably, the introduction of periodic driving profoundly reshapes this localization behavior. In addition to conventional unipolar localization, where all the eigenstates accumulate at a single corner, the driven system allows the emergence of a bipolar configuration, in which states simultaneously localize at diagonally opposite corners. A recent study has provided experimental evidence for such a unipolar-bipolar transition induced by periodic driving \cite{Zhang2026}, supporting its realization even in higher-order systems. Interestingly, under appropriate driving conditions this bipolar arrangement realizes an anomalous form of the $\mathbb{Z}_2$ skin effect \cite{Okuma2020,Okuma2023,Zhang2022,Lin2023}, a phenomenon typically associated with spinful systems possessing time-reversal symmetry and Kramers degeneracy, albeit here emerging within an effectively spinless platform.
\begin{figure}[t]
         \includegraphics[width=\columnwidth]{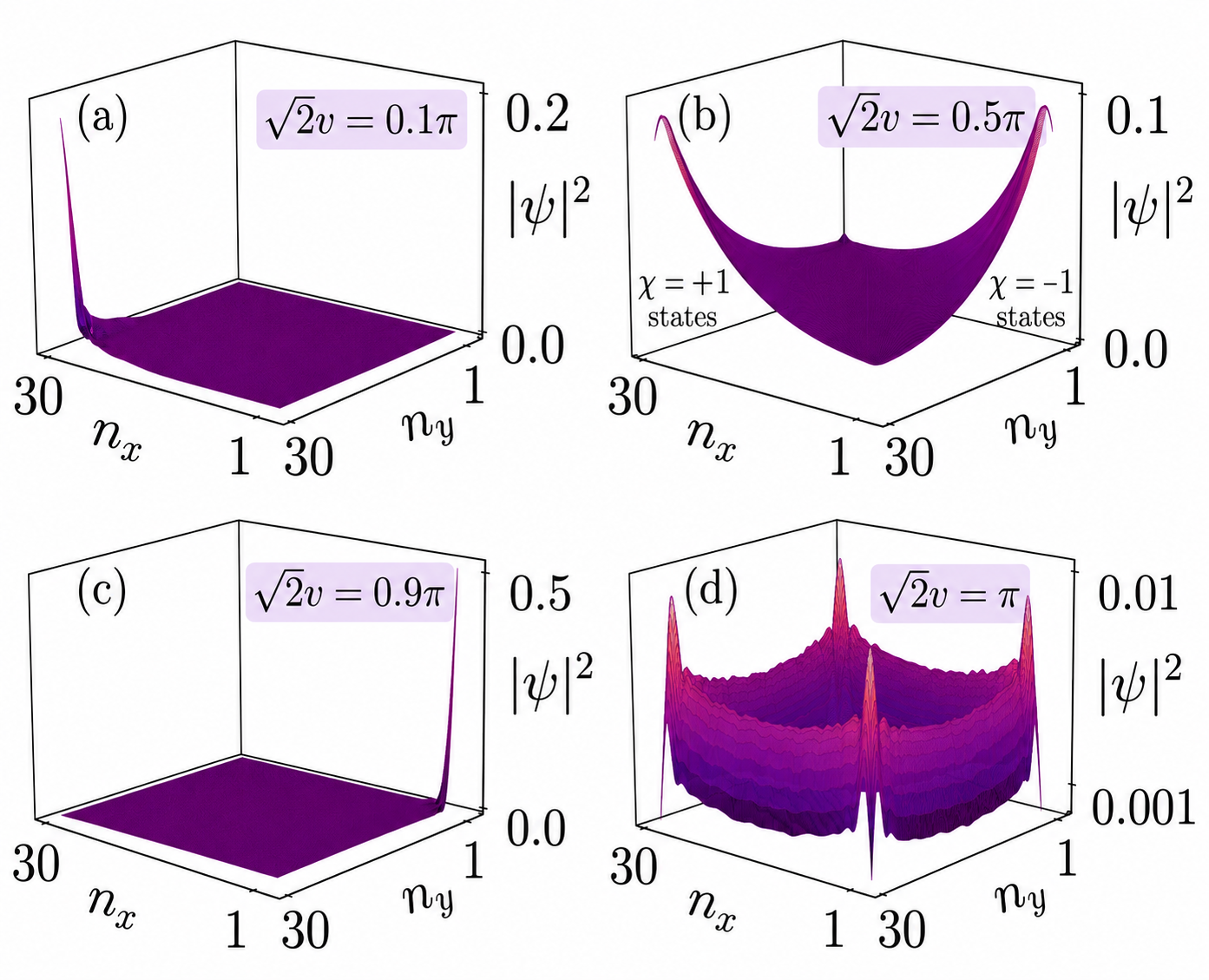}
\caption{Unipolar-bipolar localization crossover followed by a complete suppression of the skin effect is shown. Panels (a)-(d) show the spatial distribution of all eigenstates in the presence of non-reciprocity, $\gamma$. In panel (a), corresponding to $\sqrt{2} v=0.1\pi$, all eigenstates accumulate at the left corner, indicating a unipolar localization regime. In panel (b), corresponding to $\sqrt{2} v=0.5\pi$, the states are distributed equally between two of the diagonal corners; specifically, eigenstates with $\chi=1$ localize at the left corner while the $\chi=-1$ quasienergy states accumulate at the right corner, giving rise to a characteristic $\mathbb{Z}_2$-like skin effect occurring at the bidirectional points, given by $\sqrt{2} v=(2n+1)\pi/T$. In panel (c), for $\sqrt{2} v=0.9\pi$, unipolar localization reappears with reversed polarization relative to panel (a). Finally, panel (d), corresponding to $\sqrt{2} v=\pi$, shows a complete disappearance of boundary accumulation, signaling total suppression of the skin effect occurring at values, $\sqrt{2} v=2n\pi/T$. The system parameters are chosen as $T=2$, $N_x=N_y=30$, $\gamma=0.3$, and $\lambda=0.5$.
}
\label{4}
\end{figure}
\par In order to analytically characterize these transitions, we focus on the symmetric driving condition $T_1=T_2 = T/2$, which reveals a set of drive-induced special points determined by the parameters satisfying
\begin{subequations}
\begin{align}
\sqrt{2}v &= \frac{2n\pi}{T},
&& \text{skin suppression}, \label{eq:special_points_a}\\
\sqrt{2}v &= \frac{(2n+1)\pi}{T},
&& \text{bipolar/$\mathbb{Z}_2$ localization}. \label{eq:special_points_b}
\end{align}
\label{eq:special_points}
\end{subequations}
These values govern qualitative changes in spectral localization. the former corresponds to complete suppression of the skin effect, while the latter marks the onset of bipolar localization with states equally distributed between opposite boundaries. Starting from small intracell coupling, for instance $\sqrt{2}v=0.1\pi$, numerical simulations show that all bulk eigenstates accumulate near a single corner, as illustrated by the spatial probability distribution in Fig.~\ref{4}a for parameters $T_1=T_2=1$ (or $T=2)$, $\lambda=0.5$, and $\gamma=0.3$. This behavior closely mirrors the static NH limit and represents the unipolar higher-order skin phase. Note that, throughout our analysis of the non-Hermitian system, the
probability densities are calculated using the right eigenvectors of the Hamiltonian. Upon increasing $v$, eigenstates gradually redistribute toward the opposite corner, signaling the emergence of bipolar localization. At the critical value $\sqrt{2}v=0.5\pi$ (corresponding to $n=0$ in Eq. \ref{eq:special_points_b}), the system reaches a perfectly balanced configuration in which states accumulate at both the corners with equal weights. At this point, the zone center- and zone edge-quasienergy sectors localize at opposite corners, forming symmetry-linked pairs that emulate the structure of a $\mathbb{Z}_2$ skin effect. Before elaborating on the physical significance of the zone-edge and zone-centre quasienergy sectors, it is instructive to first establish the conceptual background of conventional $\mathbb{Z}_2$ localization, so as to clarify why our system does not conform to this standard paradigm and is instead more appropriately termed `$\mathbb{Z}_2$-\textit{like}'.
\par In conventional settings, a $\mathbb{Z}_2$ localization originates from spin-resolved sectors protected by the time-reversal symmetry and the Kramers degeneracy, such that the up/down spin sectors can localize at the opposite corners. In contrast, the present model contains no `real' spin degrees of freedom; instead, periodic driving dynamically generates equal numbers of quasienergy states related by the modified symmetry algebra, allowing quasienergy-opposite modes to behave as effective spin counterparts. The resulting localization therefore constitutes an \textit{anomalous $\mathbb{Z}_2$ skin effect}, demonstrating once again how Floquet-induced symmetry restructuring enables spinful phenomena to emerge in a fundamentally spinless lattice.
\par Further increasing $v$ beyond the bipolar point causes all eigenstates to migrate toward the opposite corner, as shown in Fig.~\ref{4}c for $\sqrt{2}v=0.9\pi$, thereby reversing the polarization of the higher-order skin accumulation. This localization persists until the system reaches the drive-controlled ``\emph{sweet spot}''
$\sqrt{2}v=\frac{2n\pi}{T}$,
which in our case corresponds to $\sqrt{2}v=\pi$ (for $n=1)$. At this special point, the skin effect is completely suppressed, and the bulk states assume spatially extended profiles throughout the lattice, leaving only the symmetry-protected corner modes intact. Deviating from this delocalization point reinstates the higher-order skin effect with reversed polarity. For example, states localized at the right corner for $\sqrt{2}v=0.9\pi$ relocalize at the left corner when $\sqrt{2}v=1.1\pi$ (not shown here), analogous to the configuration observed in Fig. \ref{4}a for $\sqrt{2}v=0.1\pi$. These observations reveal that periodic driving introduces a controllable delocalization–relocalization mechanism, through dynamically generated skin-suppression points. The driven system therefore exhibits a tunable, repeating sequence of unipolar, bipolar, and reversed localization regimes. Far from being a mere numerical artefact, this behaviour is a robust, symmetry-protected phenomenon, which motivates us to investigate the origin of this $\mathbb{Z}_2$-like effect and to characterize it using reliable topological indicators. 
\subsection{\label{s4B}Physical Mechanism Underlying the $\mathbb{Z}_2$-like Bipolar Skin Effect}
We start by moving along the mirror line (high-symmetry line, $k_x = k_y = k$), where the problem reduces to a two-orbital ($A$, $B$) chain driven in two
steps of duration $T_1 = T_2 = T/2$: an intracell step governed by $H_1$, and an intercell step governed by $H_2(k)$, which carries all of the momentum dependence, as well as the non-reciprocity $\gamma$. This division of labour is the key to the intuition. $H_1$ acts strictly \emph{within} a unit
cell: it contains no momentum, and therefore cannot move a particle from one cell to the next. $H_2$, by contrast, is the \emph{only} part of
the drive that can actually transport probability along the lattice.
\par Exponentiating the first step using $H_1^2 = 2v^2\mathbbm{1}$ and writing
$a \equiv \sqt\,v\,T_1 $
for the accumulated intracell phase, we obtain the exact closed form
\begin{equation}
  e^{-iH_1T_1}
  = \underbrace{\cos a\;\mathbbm{1}}_{\text{stay}}
  \;-\; \underbrace{i\,\sin a\;\sigma_x}_{\text{exchange}} .
  \label{eq:U1}
\end{equation}
A single dial, $a$, sets the balance between ``remain on the same orbital'' and ``exchange with the
other orbital''. Crucially, the exchange channel is $\sigma_x$, which is \emph{perfectly
reciprocal}: it transfers $A\!\to\!B$ and $B\!\to\!A$ with identical amplitude. At this point, it is useful to identify two important driving conditions.
\begin{enumerate}
    \item \textbf{The no-swap condition ($\sin a = 0$, i.e.\ $\sqt v = 2n\pi/T$):} Here
$e^{-iH_1T_1} = \pm\mathbbm{1}$ exactly, so the first step does nothing, and the stroboscopic dynamics
reduces to the repeated action of the second step alone. Each orbital experiences precisely the same
non-reciprocal environment in every period, with no reshuffling. As seen in \ref{4}(d), the projected accumulated bias cancels, resulting in a \emph{completely suppression} of the skin effect.
\item \textbf{The full-swap condition} ($\cos a = 0$, i.e.\ $\sqt v = (2n{+}1)\pi/T$): Here
the ``stay'' channel in Eq. \ref{eq:U1} vanishes identically and $e^{-iH_1T_1} = \mp i\,\sigma_x$: the first step
exchanges the two orbitals \emph{completely}, and reciprocally. The physical consequence lays the foundation in the sense that the drive alternates between the orbitals at which the gain and loss act.
\end{enumerate}
Therefore, note that the second step is involved applying a
non-reciprocal kick whose effect on orbital $A$ differs from its effect on orbital $B$. At the
full-swap condition, the first step swaps the two orbitals once per period, so the orbital
amplified in one period is suppressed in the next one. Combined with the reciprocal
propagation supplied by the second step, this strictly alternating pattern allows two families of
Floquet states to experience \emph{opposite} bias, which is what makes bipolar localization
possible.
\par This intuitive picture correctly identifies \emph{why} a perfect bipolar localization, that is, a $\mathbb{Z}_2$-like localization, becomes possible only at the `full-swap' points and \emph{why} it is absent at the no-swap points. It does not,
however, identify \emph{which} two families are involved, nor does it explain the exact $50{:}50$ balance. For
that we must turn to the analytical/mathematical structure supporting this intuitive idea.
\par To begin with, we note that since both step Hamiltonians are traceless, $\det U(k) = 1$ identically, and the eigenvalues of $U(k)$ satisfy $u_+ u_- = 1$ at every momentum. The entire spectral content of the $2\times2$ Floquet operator is therefore carried by a single complex function, namely the trace of the evolution operator, $\tau(k) \equiv \mathrm{Tr}\,U(k)$. Now, evaluating the trace in a closed form, within a single time-period gives
\begin{equation}
  \;
  \tau(k) \;=\; 2\cos a\,\cos b(k)
  \;-\; 2\sqt\lambda\,\sin a\,\frac{\sin b(k)}{\Lambda(k)}\,\cos k \;,
  \label{eq:trace}
\end{equation}
with $b(k) \equiv \Lambda(k)T_2$, and
\begin{equation}
    \Lambda^2(k) \equiv 2\lambda^2 - 2\gamma^2 + 4i\lambda\gamma\sin k.
    \label{eq:lambda}
\end{equation}
The two special driving conditions of Eq.~\eqref{eq:special_points} acquire a transparent geometric meaning once we examine the mathematical structure of the trace. First, note that under a shift by half the Brillouin zone (using $\sin(k+\pi) = -\sin k$, with $\lambda,\gamma$ real), the trace obeys
\begin{equation}
  \Lambda^2(k+\pi) = 2\lambda^2-2\gamma^2-4i\lambda\gamma\sin k = \big[\Lambda^2(k)\big]^{*} .
\end{equation}
Since $\cos b$ and $\sin b/\Lambda$ are functions of $\Lambda^2$ with \emph{real} coefficients of the Taylor expansion, this conjugation property can be carried over directly to Eq.~\eqref{eq:trace}. Setting $\cos(a) = 0$ at the full-swap (perfect bipolar) condition then yields the exact identity
\begin{equation}
  \tau(k+\pi) = -\,\big[\tau(k)\big]^{*}\,,
  \label{eq:antipodal}
\end{equation}
valid for every $k$ at the full-swap condition.
\par Let us now understand the consequence of this special feature at the level of the quasienergy spectra. Recall that the eigenvalues obey, 
\begin{equation}
    u^2 - \tau(k)u + 1 = 0.
\end{equation} 
Taking
the complex conjugate of this equation gives $(u^*)^2 - \tau(k)^*u^* + 1 = 0$, and substituting
$w = -u^*$ yields $w^2 + \tau(k)^*w + 1 = 0$, that is $w^2 - \tau(k+\pi)w + 1 = 0$ by
Eq.~\eqref{eq:antipodal}. Hence $-u^*$ solves the characteristic equation at momentum $k+\pi$, and the two sets of eigenvalues are related via,
\begin{equation}
  \big\{u_\pm(k+\pi)\big\} = \big\{-\,u_\pm(k)^{*}\big\},
  \label{eq:umap}
\end{equation}
Writing $u = e^{-i E T}$,
the map $u \mapsto -u^{*}$ acts on the quasienergy as,
\begin{equation}
  \;E \;\longmapsto\; -\,E^{*} - \frac{\pi}{T}\;
  \label{eq:epsmap}
\end{equation}
which sends $0 \mapsto -\pi/T \equiv \pi/T$ and $\pi/T \mapsto 0$. This mapping is an involution: applying $u \mapsto -u^{*}$ twice returns $-(-u^{*})^{*} = u$, and shifting the momentum twice by $\pi$ returns $k$. We therefore define, \begin{equation}
  \Theta : (k,u) \;\longmapsto\; (k+\pi,\,-u^{*}),
  \qquad \Theta^2 = \mathbbm{1},
  \label{eq:Theta}
\end{equation}
As a consequence, $\Theta$ partitions the Floquet zone into two arcs. The two fixed points
$E = \pm\pi/2T$ divide the quasienergy circle into two arcs, which $\Theta$ exchanges:
the arc containing the zone centre and the arc containing the zone edge. The corresponding binary
label is
\begin{equation}
  \;
  \chi \;\equiv\; \mathrm{sgn}\big(\mathrm{Re}\,u\big)
  \;=;
  \begin{cases}
    +1, & |\mathrm{Re}\,E| < \pi/2T \quad(\text{zone-centre}),\\[2pt]
    -1, & |\mathrm{Re}\,E| > \pi/2T \quad(\text{zone-edge}),
  \end{cases}
  \;
  \label{eq:chi}
\end{equation}
and under $\Theta$ one has $\mathrm{Re}\,u \to -\mathrm{Re}\,u$, resulting in $\chi \to -\chi$ exactly.
This is the spinless analogue of the Kramers pairing. In the conventional spinful case
TRS$^\dagger$ pairs the spin-$\uparrow$ and spin-$\downarrow$ partners to
opposite edges. Here the involution $\Theta$ pairs $E$ with $-E-\pi/T$ and sends
the \emph{zone-centre family} ($\chi=+1$) and the \emph{zone-edge family} ($\chi=-1$) to opposite
corners.
\subsection{Symmetry Protection of the $\mathbb{Z}_2$-like Effect}
The spectral involution $\Theta$ introduced in Sec.~\ref{s4B} is not a mere numerical coincidence but the joint consequence of three exact symmetries of the driven lattice. First, the chiral symmetry $S = \sigma_z$, under which both step Hamiltonians anticommute with the sublattice grading, satisfies $S\,U(k)\,S^{-1} = U(k)^{-1}$ and thus inverts the eigenvalue, $u \to 1/u$, or equivalently $E \to -E$. Second, TRS$^\dagger$ acts as a transpose combined with $k \to -k$; since 
\begin{equation}
    H_1 = H_1^{\mathsf{T}} \quad \text{and} \quad  H_2(k+\pi) = -\big[H_2(-k)\big]^{\mathsf{T}},
\end{equation} 
this operation interchanges the left- and right-going hopping amplitudes and thereby reverses the direction of non-reciprocal amplification, effectively flipping the corner at which the skin mode accumulates. Third, at the full-swap (bipolar) driving condition the first half-period reduces to a pure exchange of the two orbitals, $e^{-iH_1T_1} = -i\sigma_x$, which squares to $-\mathbbm{1}$; this supplies an overall factor $u \to -u$, i.e., a shift of the quasienergy by half the Floquet zone, $E \to E + \pi/T$, and is what ultimately exchanges the zone-centre and zone-edge branches. Carrying these three identities through the one-period evolution $U(k) = e^{-iH_2(k)T_2}e^{-iH_1T_1}$ combines them into a single relation, 
\begin{equation}
    U(k+\pi) = -\big[U(-k)^{\mathsf{T}}\big]^{-1},
\end{equation} where the transpose encodes TRS$^\dagger$, the inverse encodes chiral symmetry, and the overall sign encodes the ful orbital exchange. At the level of eigenvalues this reads
\begin{equation}
    u(k+\pi) = -u(k)^{*},
\end{equation} which is precisely the action of $\Theta:(k,u)\mapsto(k+\pi,-u^{*})$. Physically, the three symmetries act in succession on a Floquet state: chiral symmetry inverts the eigenvalue, the orbital exchange shifts the quasienergy by half the zone and carries a zone-centre state into the zone-edge family, and TRS$^\dagger$ finally flips the corner, so that $|\text{zone-centre, left corner}\rangle$ is mapped to $|\text{zone-edge, right corner}\rangle$, which can be held precisely as the spinless analogue of $|\!\uparrow,\text{left}\rangle \to |\!\downarrow,\text{right}\rangle$ in the conventional spinful $\mathbb{Z}_2$ skin effect, with the Floquet sector label $\chi$ playing the role of spin.
\begin{figure*}[t]
\centering
\includegraphics[width=0.8\textwidth]{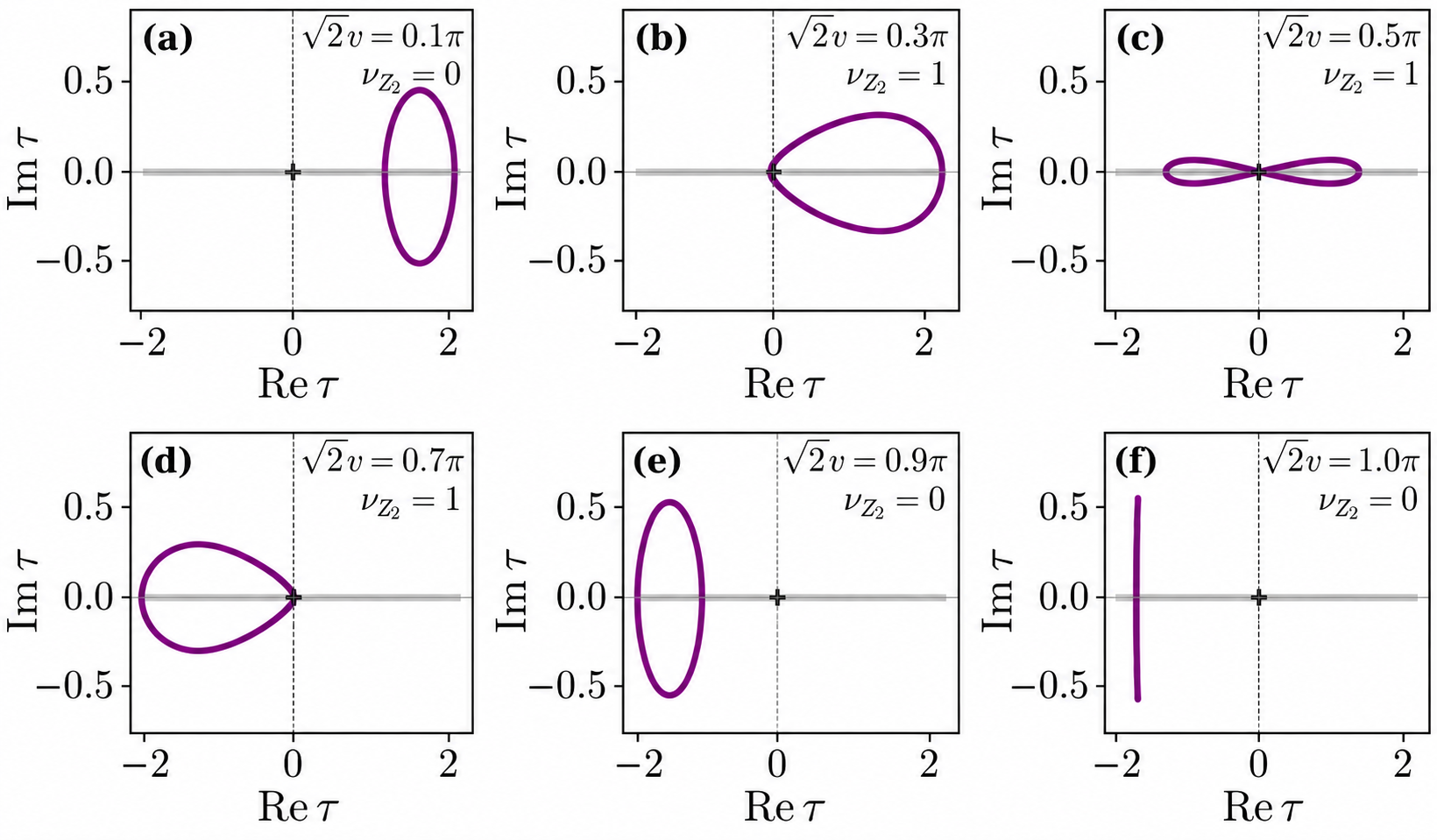}
\caption{Trace trajectories $\tau(k) = \mathrm{Tr}\,U(k)$ in the complex plane as the crystal momentum $k$ sweeps the Brillouin zone, for driving amplitudes $\sqrt{2}v = 0.1\pi$ (a), $0.3\pi$ (b), $0.5\pi$ (c), $0.7\pi$ (d), $0.9\pi$ (e), and $1.0\pi$ (f), with $\gamma=0.3$, $\lambda=0.5$, and $T=2$. The black cross denotes the origin, while the grey band indicates the real axis. The Floquet $\mathbb{Z}_2$ bipolar invariant, $\nu_{\mathbb{Z}_2} = \tfrac12[1 - \mathrm{sgn}(\tau(0)\tau(\pi))]$, identifies whether the trace loop encloses the origin: $\nu_{\mathbb{Z}_2} = 1$ when the origin lies inside the loop [panels (b)--(d)], corresponding to the bipolar localization regime, and $\nu_{\mathbb{Z}_2} = 0$ when it lies outside [panels (a) and (e)], corresponding to the unipolar regime. At the perfect bipolar point (c), the trace satisfies $\tau(0) = -\tau(\pi)$ and is symmetric about the origin, whereas at the suppression point (f), the trace trajectory retraces itself, encloses zero area, and no longer exhibits spectral winding.}
\label{trace_trajectory}
\end{figure*}
\subsection{The Floquet version of $\mathbb{Z}_2$ invariant: Trace-Winding Index}
Since the effect identified above is symmetry-enforced rather than accidental, an associated topological invariant must exist, which we shall construct in the following. The conventional routes, however, are unavailable. For an ordinary (unipolar) skin effect in a spinless setting, the standard diagnostic is the $\mathbb{Z}$-valued spectral winding number ($W$) \cite{Okuma2020}, with a nonzero winding accompanied by a generalized Brillouin zone (GBZ) \cite{Yokomizo2019,Zhang2020,Wu2020} of radius $\neq 1$. In the spinful $\mathbb{Z}_2$ case, this diagnostic instead fails outright. The total winding vanishes, $W_\uparrow + W_\downarrow = 0$, the GBZ collapses onto the unit circle, and a $\mathbb{Z}_2$ index is extracted only by splitting the characteristic polynomial (det [$H(\beta)-E]=0$, where $\beta = e^{ik}$ defines the contour of GBZ) which solves for the GBZ into two decoupled sectors --- one furnishing a sub-GBZ of radius $>1$ and the other of radius $<1$ --- and taking the winding of a single sector as modulo two. However, neither of the routes apply here: because $\Lambda(k)$ of Eq.~\eqref{eq:lambda} is momentum-dependent, the characteristic equation involves the transcendental functions $\cos b(k)$ and $\sin b(k)/\Lambda(k)$, and cannot be split into two reciprocal sub-equations (a more detailed discussion on the characteristic polynomial and evaluation of the GBZ is provided in the next section). What is needed instead is a construction that captures the two graded families without requiring the polynomial to factorize, and the trace of the Floquet evolution operator introduced in Sec. \ref{s4B} supplies exactly this.
\begin{table}[t]
\centering
\footnotesize
\renewcommand{\arraystretch}{1.3}
\begin{tabular}{cccccl}
\toprule
$\sqrt{2}v$ & $\tau(0)$ & $\tau(\pi)$ & $\tau_0$ & $\nu_{\mathbb{Z}_2}$ & regime \\
\midrule
$0.100\pi$ & $+1.192$ & $+2.020$ & $+1.606$ & $0$ & deep unipolar (Fig. \ref{4}a) \\
$0.300\pi$ & $-0.092$ & $+2.077$ & $+0.992$ & $1$ & bipolar, imbalanced \\
$\star\;0.500\pi$ & $-1.340$ & $+1.340$ & $0.000$ & $1$ & $\mathbb{Z}_2$-like bipolar (Fig. \ref{4}b) \\
$0.700\pi$ & $-2.077$ & $+0.092$ & $-0.992$ & $1$ & bipolar, reversed skew \\
$0.900\pi$ & $-2.020$ & $-1.192$ & $-1.606$ & $0$ & unipolar, reversed (Fig. \ref{4}c) \\
$\star\;1.000\pi$ & $-1.688$ & $-1.688$ & $-1.688$ & $0$ & skin suppression (Fig. \ref{4}d) \\
\bottomrule
\end{tabular}
\caption{Trace intercepts $\tau(0)$, $\tau(\pi)$, midpoint $\tau_0 = 1/2 [ \tau(0) + \tau(\pi) ]$, and winding index $\nu_{\mathbb{Z}_2}$ across the driving range of Fig.~\ref{trace_trajectory}, for $\gamma=0.3$, $\lambda=0.5$, $T=2$. Rows marked $\star$ are the exact special points of Eq.~\eqref{eq:special_points}.}
\label{tab:winding}
\end{table}
Note that, for a $2\times 2$ Floquet operator the characteristic polynomial collapses onto the trace, $\det[U(k)-u] = u[\zeta - \tau(k)]$, with $\zeta = u + u^{-1} = 2\cos(E T)$. The spectral winding of the Floquet spectrum around any reference eigenvalue $u$ on the unit circle is therefore equivalently the winding of the trace curve around the corresponding real point $\zeta = 2\cos(E T) \in [-2,2]$,
\begin{equation}
  W(\zeta) = \frac{1}{2\pi i}\oint_{\rm BZ} dk\;\partial_k \ln\big[\tau(k) - \zeta\big] \in \mathbb{Z}.
  \label{eq:W}
\end{equation}
This index is quantized and gauge-independent; it treats the two Floquet branches (both zone center and zone edge) jointly, since $\tau = u_+ + u_-$, so that no factorization of the characteristic polynomial is required. A nonzero $W(\zeta)$ signals skin accumulation of the states with quasienergies near $\zeta$, with the sign of $W$ selecting the corner --- the direct analogue of the winding--skin correspondence of static non-Bloch theory \cite{Okuma2020}. By Eq.~\eqref{eq:antipodal}, the map $\tau \to -\tau^{*}$ reverses the orientation of the trace curve, so that at the full-swap point $W(-\zeta) = -W(\zeta)$. The total winding across the sector boundary therefore vanishes identically even though each sector individually winds (the structural analogue of $W_\uparrow = -W_\downarrow$), and the precise reason for a $\mathbb{Z}_2$, rather than a $\mathbb{Z}$, index is necessitated.

From the discussion in Sec. \ref{s4B}, the trace curve crosses the real axis at $k=0,\pi$; when these are the only interior crossings, $W(\zeta)$ can change only across the interval bounded by $\tau(0)$ and $\tau(\pi)$, and the surviving index reduces to the closed parity form
\begin{equation}
  \nu_{\mathbb{Z}_2} = \tfrac12\Big[\,1 - \mathrm{sgn}\big(\tau(0)\,\tau(\pi)\big)\Big] \in \{0,1\},
  \label{eq:nu}
\end{equation}
in which $k=0$ and $k=\pi$ play the role of the time-reversal-invariant momenta of the Fu--Kane construction. We emphasize the logical status of this formula: $\nu_{\mathbb{Z}_2}=1$ exactly when the two real intercepts straddle a sector of nonzero winding on each side of $\zeta = 0$, i.e., exactly when the two graded families accumulate at the opposite corners.

Table~\ref{tab:winding} lists $\tau(0)$, $\tau(\pi)$, the intercept midpoint $\tau_0$, and the resulting index $\nu_{\mathbb{Z}_2}$ across the driving range of Fig.~\ref{4}, confirming that the evolution of the index reproduces the full phenomenology of the localization transition.
Moreover, the evolution of the index, as shown in Fig. \ref{trace_trajectory} reproduces the phenomenology of Fig.~\ref{4} in complete detail, and
clarifies the unipolar-bipolar transition in the following way. At small driving ($\sqrt{2}v = 0.1\pi$) all the states accumulate at a single corner, the deep unipolar regime of Fig.~\ref{4}(a). On increasing the drive ($\sqrt{2}v = 0.3\pi$), states begin to appear at the opposite corner and the system becomes bipolar; however, this is bipolar without yet being $\mathbb{Z}_2$, since the two intercepts remain highly unequal in magnitude ($|{-}0.092|$ versus $|{+}2.077|$), so the two corners are populated with strongly unequal weight. At the full-swap condition ($\sqrt{2}v = 0.5\pi$), the index remains finite, $\nu_{\mathbb{Z}_2} = 1$, but the two intercepts are now exactly equal and opposite, $\tau(0) = -\tau(\pi)$, or equivalently $\tau_0 = 0$. This is the signature of genuine $\mathbb{Z}_2$-like behaviour: the vanishing of $\tau_0 = \tfrac{1}{2}[\tau(0)+\tau(\pi)]$ is precisely the condition $\tau(k+\pi) = -\tau(k)$ of Eq.~\eqref{eq:antipodal}, forcing the $\chi=+1$ (zone-centre) and $\chi=-1$ (zone-edge) families onto opposite corners with strictly equal weight, corresponding to the balanced bipolar distribution of Fig.~\ref{4}(b). Beyond this point ($\sqrt{2}v = 0.7\pi$), the index remains finite but the imbalance returns with opposite skewness, and by $\sqrt{2}v = 0.9\pi$, the intercepts again share a common sign, so that $\nu_{\mathbb{Z}_2}$ returns to zero: unipolar localization with reversed polarity, as seen in Fig.~\ref{4}(c). Finally, at the suppression point ($\sqrt{2}v = 1.0\pi$), the no-swap condition $\sin(a)=0$ eliminates the $k$-odd term in the trace, leaving $\tau(k) = \pm 2\cos b(k)$. Therefore, the trace curve retraces itself and encloses zero area, so the spectral winding vanishes for every reference point $\zeta$ and no Floquet eigenvalue winds around the unit circle. Consequently, the non-Hermitian skin effect is completely suppressed and the bulk spectrum remains extended, as reported in Fig.~\ref{4}(d).

\subsection{\label{s5}Non-Bloch characterization via Floquet GBZ}
The preceding sections have provided a comprehensive understanding of the physical origin of higher-order unipolar–bipolar skin localization, together with symmetry-resolved approaches for characterizing the associated skin modes. A natural question that arises at this stage is how the topological states behave in the same regime, particularly since they can also become localized at the corners. We therefore devote this section to a detailed investigation of these corner-localized topological states and their underlying mechanism. Since the driven lattice operates in the NH regime and supports pronounced skin accumulation, conventional Bloch-band invariants are no longer adequate for identifying the topological origin of the boundary states. In particular, the presence of the NHSE invalidates the standard BBC strictly defined within the Bloch framework. A consistent topological description therefore, requires the non-Bloch formalism, where the topology is defined on a generalized Brillouin zone (GBZ) \cite{Yokomizo2019,Zhang2020,Wu2020}. The central idea of this approach is to reformulate the Bloch Hamiltonian via replacing the momentum variable, $k$, with a phase factor, namely, $\beta=e^{ik}, (k \in [-\pi, \pi])$. The Hamiltonian can then be expressed as a non-Bloch operator $H(\beta)$. The allowed values of $\beta$ are determined from the characteristic equation $\det [H(\beta) - E] = 0$. By arranging the resulting solutions according to their absolute values and selecting those compatible with the open-boundary condition, one obtains a closed contour in the complex $\beta$-plane that defines the GBZ. This contour provides the correct bulk description for systems exhibiting the NHSE.
\par While the non-Bloch method has been successfully applied to a wide range of one-dimensional NH systems \cite{Yokomizo2021,Roy2025Prime,Roy2025DoublePrime,Zhang2020Prime,Wu2020}, extending it to two-dimensional lattices poses significant complications. The presence of two independent momentum components generally produces characteristic equations involving multiple complex variables, making the analytical construction of the GBZ highly cumbersome. To simplify the analysis, we take advantage of the mirror symmetry introduced in Eq. \ref{eq:mirror_symmetry}, which also plays a central role in the emergence of higher-order topology in the model. In particular, we restrict our attention to the high-symmetry line, $k_x=k_y=k$ \cite{Liu2019}, which effectively reduces the two-dimensional Bloch Hamiltonian to a block off-diagonal form,
\begin{equation}
\tilde{H} =
\begin{pmatrix}
H^{+} & 0 \\
0 & H^{-}
\label{eq:Ham_in_symmetry_line}
\end{pmatrix}.
\end{equation}
where each block $H^\alpha (\alpha=\pm)$ takes the form of a massless Dirac Hamiltonian. When written in this representation, the two-component structure naturally allows one to define a winding number for each block. In this context, the higher-order topology of the system can be characterized by the mirror-graded winding number defined as,
\begin{equation}
\nu = \frac{\nu^{+} - \nu^{-}}{2},
\label{eq:mirror_graded_winding}
\end{equation}
where $\nu^{\alpha}$ ($\alpha=\pm$) denotes the conventional winding number associated with each block $H_{\alpha}$ (details are provided in Appendix. \ref{app2}).
\par
\begin{figure}[t]
         \includegraphics[width=\columnwidth]{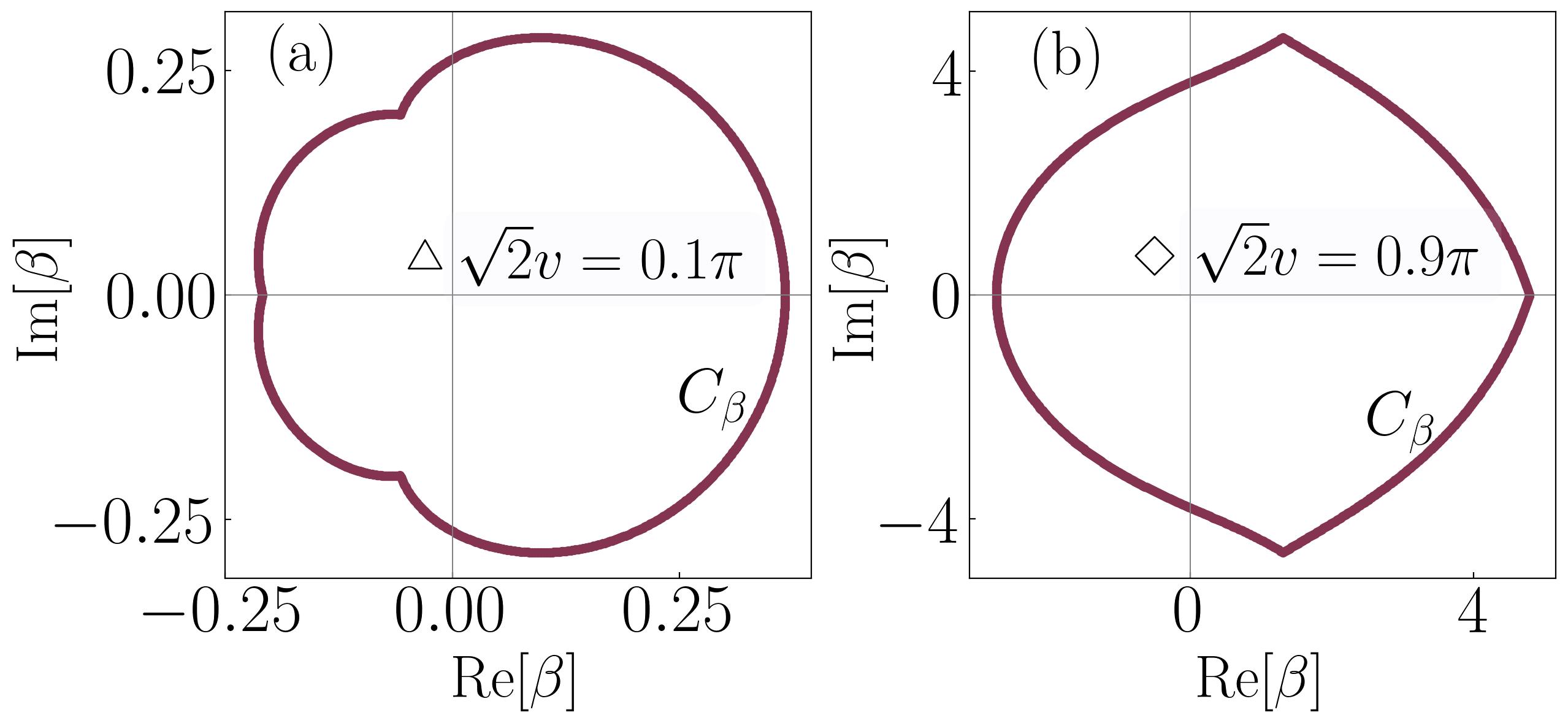}
\caption{Closed contours of the 1D mapped Floquet GBZ along the symmetry line $k_x = k_y$, plotted in the complex plane $\mathrm{Re}(\beta)$ versus $\mathrm{Im}(\beta)$. Panel (a), corresponding to $\sqrt{2}v=0.1\pi$, lies in the unipolar regime where all the eigenstates accumulate at the left boundary; the associated GBZ forms a closed loop with radius $|\beta|<1$. In contrast, panel (b), for $\sqrt{2}v=0.9\pi$, corresponds to right-boundary accumulation, and the GBZ contour has radius, $|\beta|>1$. In both the cases, the contours exhibit cusp-like behavior, reflecting the multiple solutions (more than two) of $\beta$ having similar amplitude. This turns out to be a direct consequence of the second-order Taylor expansion of the characteristic polynomial governing the Floquet operator. The system parameters are chosen as $T=2$, $\lambda=0.5$, and $\gamma=0.3$.
}
\label{5}
\end{figure}

 Nevertheless, extending this problem to a periodically driven scenario introduces further subtleties that prevent the straightforward application of this procedure. Unlike the static Hamiltonians, which typically contain only short-range interactions and therefore yield characteristic polynomials of finite order in $\beta$, the effective Hamiltonian generated by periodic driving can involve longer-range interactions. This can be seen directly from the momentum-space representation of the driven Hamiltonian (see Appendix. \ref{app2}), where each $ d$-vector contains functions such as $f[\cos(\sin k)]$. Consequently, the associated characteristic equation may become infinite order in $\beta$. Furthermore, the Floquet effective Hamiltonian generally does not retain the block-diagonal structure along the high-symmetry line, which prevents the direct definition of the mirror-graded invariant. To overcome this issue, we return to the symmetric time-frame formulation formed via the unitary operation (See Eq. \ref{eq:symmetric_frame}) discussed in Sec. \ref{s3}. As demonstrated in Appendix. \ref{app2}, expressing the Floquet operator in these symmetric frames restores the massless Dirac structure along the high-symmetry line, thereby recovering the block decomposition required for defining the mirror-graded invariant. In other words, the topological invariant becomes well defined only when evaluated within these symmetric Floquet representations. The remaining challenge lies in reducing the infinite-order characteristic equation to a tractable form. To this end, we perform a Taylor expansion of the characteristic polynomial which requires rewriting the characteristic equation in the following way,
\begin{equation}
    E_{\alpha}^2(\beta) = d_x^2(\beta) + (\alpha)^2 d_y^2(\beta).
    \label{eq:characteristic}
\end{equation}
The exact expressions for the $d$-vectors are provided in Appendix. \ref{app2}. Thus, considering a small $\lambda$ (say $\lambda<1$) allow us performing a Taylor expansion of Eq. \ref{eq:characteristic} upto second order with respect to $\lambda$. Consequently, the characteristic equation corresponding to each block in Eq. \ref{eq:Ham_in_symmetry_line} can be expressed as,
\begin{equation}
E_{\alpha}^2(\beta) = \sum_{j=-2}^{2} \beta^j X_{j},
\label{Taylor}
\end{equation}
where $X_{j}$ are coefficients associated with powers of $\beta$. Also note that, the degree of the polynomial depends on the order at which the Taylor expansion is truncated. 
Now assume there are two solutions $\beta$ and $\beta' = \beta e^{i\theta}$ that share the same magnitude. Subtracting the two characteristic equations $E^{2} = F(\beta)$ and $E^{2} = F(\beta')$ yields
\begin{equation}
0 = \sum_{j=-2}^{2} \beta^j X_{j} (1 - e^{ij\theta}),
\label{Taylor2}
\end{equation}
which allows us to determine the solutions for $\beta$ over the range $\theta \in [0,2\pi]$.
Finally, the $\beta$ solutions are arranged in ascending order based on their absolute values. By selecting those that satisfy the condition
\begin{equation}
|\beta_{2N}| = |\beta_{3N}|,
\end{equation}
with $N$ denoting the number of discretization intervals in $\theta$, we obtain the coordinates of GBZ, $C_{\beta}$. Interestingly, although this construction represents a one-dimensional projection of the full two-dimensional GBZ, it remains sufficient to determine the localization direction of the skin modes. For example, when all the eigenstates accumulate at the left corner e.g., $\sqrt{2}v = 0.1\pi$, the resulting GBZ has a radius smaller than 1 (see Fig. \ref{5}a), whereas localization at the right corner for $\sqrt{2}v = 0.9\pi$,  corresponds to a GBZ with radius larger than 1 (see Fig. \ref{5}b). In general, the GBZ deviates from the conventional Brillouin zone, reflecting the non-Bloch nature of the spectrum. Additionally, the GBZs obtained from the two symmetric time frames coincide in both shape and size, validating the accuracy of the truncation scheme.
\begin{figure}[t]
         \includegraphics[width=1\columnwidth]{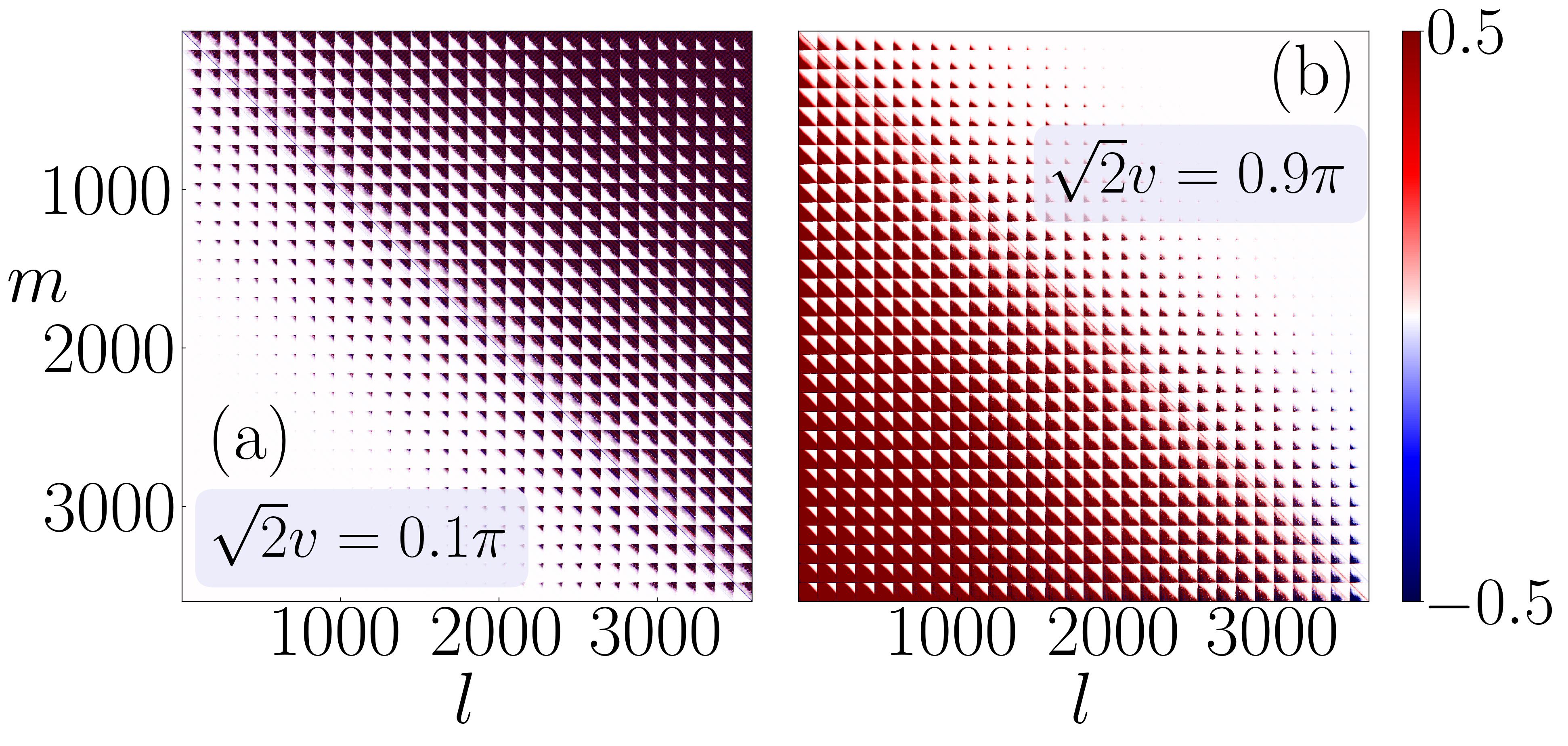}
\caption{Expansion coefficients of the driven effective Hamiltonian $H_{\mathrm{eff}}$ in the coordinate basis for $\sqrt{2}v = 0.1\pi$ and $\sqrt{2}v = 0.9\pi$, shown in panels (a) and (b), respectively. Here, $l$ and $m$ denote the basis indices corresponding to the row and column positions of the Hamiltonian matrix. For $\sqrt{2}v = 0.1\pi$, where all wavefunctions are localized at the left corner, the upper block of the matrix becomes populated (with the largest amplitudes concentrated near the corner), while the lower block remains essentially empty. Conversely, for $\sqrt{2}v = 0.9\pi$, where the wavefunctions localize at the right corner, the lower block becomes populated while the upper block remains empty. The system parameters are $N_x = N_y = 30$, $T= 2$, $\lambda = 0.5$, and $\gamma = 0.3$.}
\label{6}
\end{figure}
\par Interestingly, the Floquet GBZ not only provides information that determines the directional accumulation of skin modes, but the drive-induced longer-range couplings also offer insight into the specific corner at which the wavefunctions accumulate. To gain further insight into this behavior, it is useful to examine the matrix configuration of the effective Hamiltonian, $\mathcal{H_{\text{eff}}}$. Starting with the Hermitian limit, expanding $\mathcal{H_{\text{eff}}}$ using the Baker–Campbell–Hausdorff series shows that increasing the driving period $T$ progressively generates higher-order commutator terms. These additional contributions effectively introduce longer-range hopping processes, causing off-diagonal matrix elements to emerge and gradually populate regions of the Hamiltonian that remain empty in the static limit. Even in the low-frequency regime, the effective Hamiltonian may acquire a dense structure in which nearly all the matrix elements become populated. At this stage, the inclusion of the non-reciprocity leads to a striking asymmetry in the matrix structure. For instance, corresponding to certain parameter regimes, either the upper or the lower block of the Hamiltonian becomes fully populated while the opposite block remains largely empty. This asymmetry directly determines the specific corner at which the wavefunctions accumulate. For example, Fig.~\ref{6}a corresponds to the parameter value $\sqrt{2}v = 0.1 \pi$, where all the eigenstates localize at the left corner. The associated matrix configuration clearly shows that the upper block is densely filled while the lower block remains essentially empty. Moreover, the largest matrix amplitudes concentrate near the corner entries, further reinforcing the emergence of the drive-induced second-order skin effect. In contrast, when $\sqrt{2}v = 0.9\pi$ (Fig.~\ref{6}b), the situation is reversed. The lower block of the Hamiltonian becomes populated while the upper block is nearly empty. The corresponding spatial distribution shows that the wavefunctions now accumulate at the right corner. These results establish a clear connection between the block structure of the effective Hamiltonian and the direction of higher-order skin localization.
\begin{figure}[t]
         \includegraphics[width=\columnwidth]{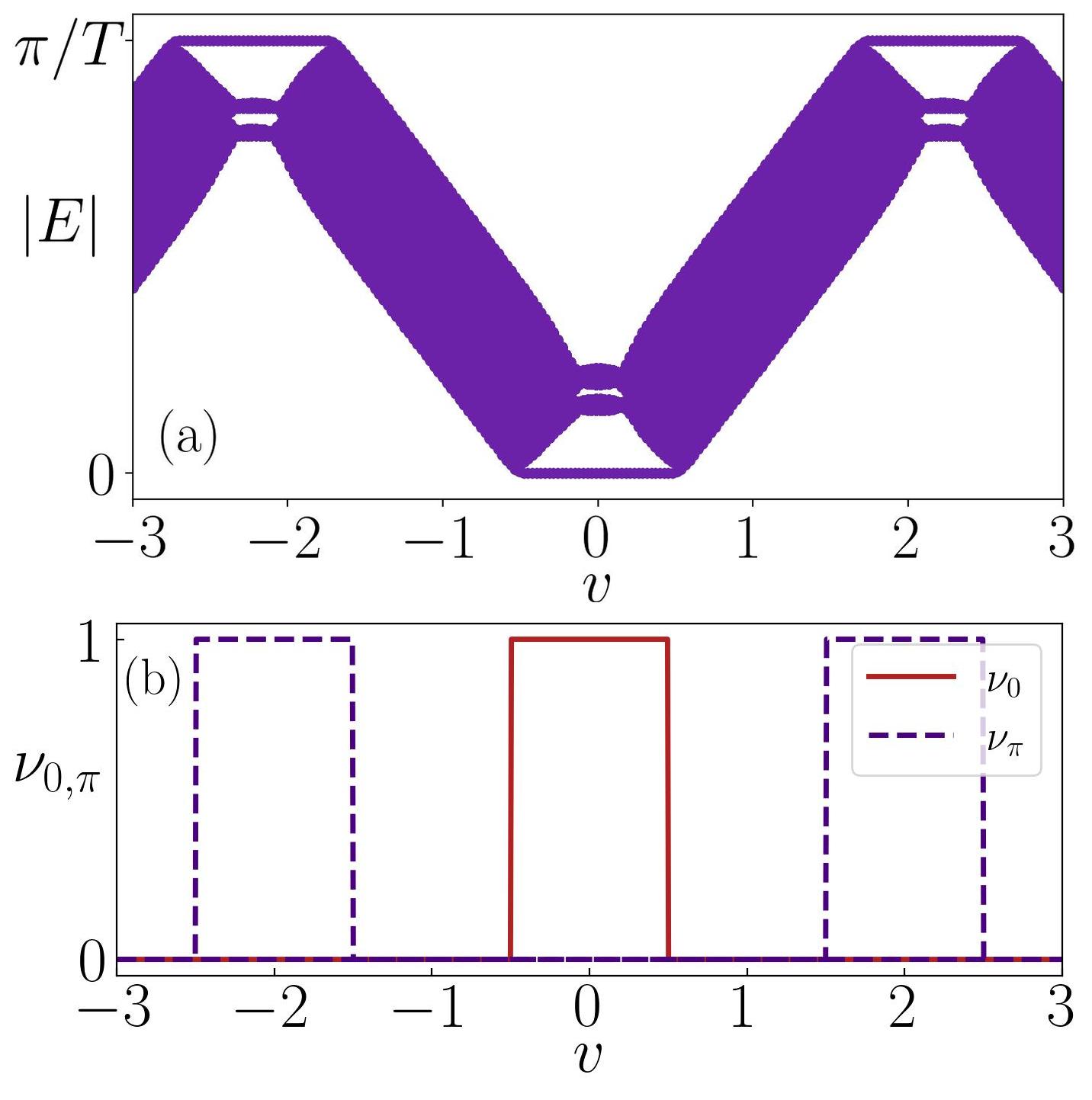}
\caption{Panel (a) shows the quasienergy spectrum under fully open boundary conditions along both the $x$- and $y$-directions, plotted as a function of $v$. The spectrum clearly exhibits the emergence of localized corner states at both zero and $\pi/T$ quasienergies. Panel (b) presents the mirror-graded winding numbers $(\nu^{0}, \nu^{\pi})$ as a function of $v$, which show precise correspondence with the boundary spectrum. Finite values of these invariants correctly signal the appearance of corner states, thereby restoring the bulk-boundary correspondence in the driven NH system. The parameters are chosen as $T=2$, $\lambda=0.5$, $\gamma=0.3$, and $N_x=N_y=30$.
}
\label{7}
\end{figure}
\par Once the GBZ is constructed, the winding numbers corresponding to each block Hamiltonian $H^{\alpha}$ can be computed by performing the contour integral over $C_{\beta}$ instead of the conventional Bloch contour $C_k$. Following the definition shown in Eq. \ref{eq:mirror_graded_winding}, we obtain the mirror-graded winding number. Extending this idea to both symmetric time frames yields two invariants, $\nu_1$ and $\nu_2$ corresponding to the operators $U_1$ and $U_2$ (which were obtained performing the transformation as shown in Eq. \ref{eq:symmetric_frame}). While each invariant alone cannot distinguish the different localized states, the Floquet classification \cite{Roy2017} allows the construction of a pair of physically meaningful invariants \cite{Asboth2013,Asboth2014} (analogous to Eq. \eqref{eq:pair}),
\begin{equation}
\nu_0 = \frac{\nu_1 + \nu_2}{2}, \qquad 
\nu_{\pi} = \frac{\nu_1 - \nu_2}{2},
\label{eq:winding_two_frames}
\end{equation}
which characterize emergence of the corner states at quasienergies $0$ and $\pi/T$, respectively. To verify this framework, we compute the quasienergy spectrum under fully open boundary conditions. As shown in Fig.~\ref{7}(a), corner-localized modes appear simultaneously at quasienergies $0$ and $\pi/T$ as the parameter $v$ is varied. While the conventional mirror-graded winding number fails to capture the onset of these states, the Taylor-expanded non-Bloch invariants $\nu_{0}$ and $\nu_{\pi}$ correctly reproduce their emergence. The evolution of these invariants as a function of $v$, shown in Fig.~\ref{7}(b), exhibits excellent agreement with the quasienergy spectrum, thereby restoring the BBC in the periodically driven NH system. 
\begin{figure}[t]
         \includegraphics[width=\columnwidth]{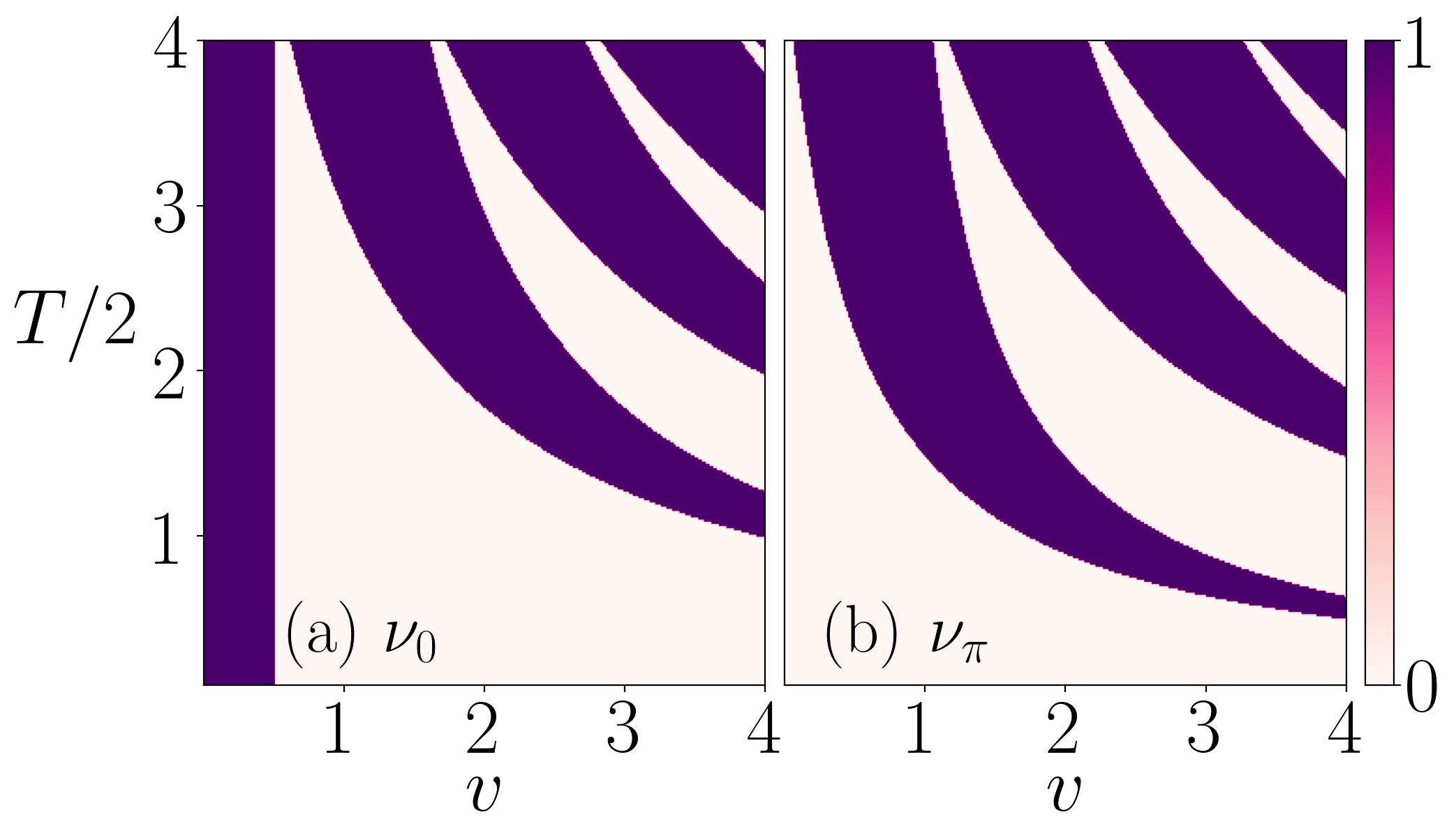}
\caption{Topological phase diagram in the $v-T/2$ plane computed using the mirror-graded winding numbers. Panel (a) shows the invariant $\nu^{0}$ corresponding to corner states at zero quasienergy, while panel (b) displays $\nu^{\pi}$ associated with corner states at quasienergy $\pi/T$. The system parameters are chosen as $\lambda = 0.5$ and $\gamma = 0.3$.}
\label{8}
\end{figure}
\par Furthermore, to provide a comprehensive overview of the distinct nontrivial phases realized in the system, we present the topological phase diagram in the $v-T/2$ plane corresponding to corner states at both zero and $\pi/T$ quasienergies, as shown in Fig.~\ref{8}. The phase diagram clearly identifies the regions where the system supports different Floquet topological phases characterized by the emergence of localized corner modes in distinct quasienergy gaps. In particular, for sufficiently large values of $v$ and the driving period $T$, the system enters a robust nontrivial regime where corner states appear simultaneously at both zero and $\pi/T$ quasienergies. This indicates the coexistence of multiple topological sectors over an extended parameter range, further emphasizing the rich topological landscape of the driven system and demonstrating how periodic driving can generate hybrid higher-order topological phases. In this way, the non-Bloch Floquet formalism provides a unified framework for capturing both the higher-order topology and the directionality of skin localization in the driven NH BBH lattice.
\begin{figure}[t]
\centering
\includegraphics[width=0.85\columnwidth]{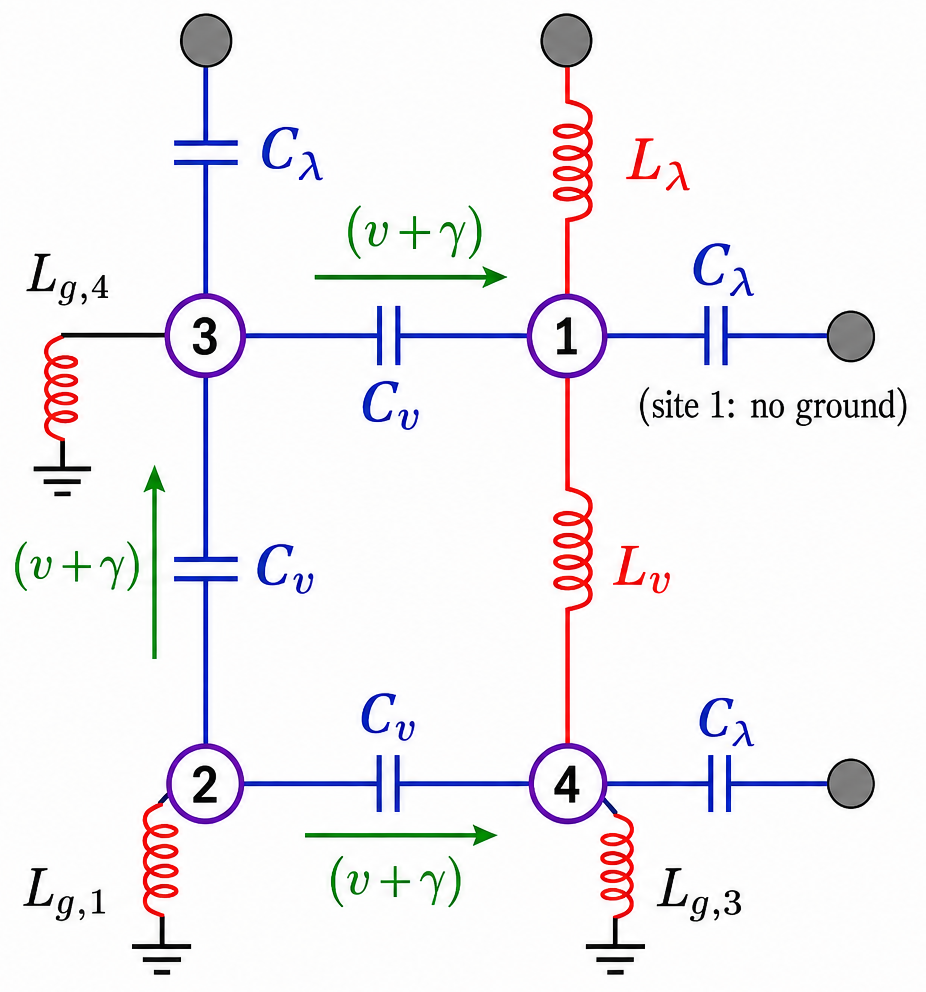}
\caption{Circuit unit cell of the NH BBH TEC. Nodes $1,3,4,2$ are the four sublattices. Each intracell bond carries a capacitor (or inductor, on the $\pi$-flux bond) in parallel with an INIC whose amplified direction (green) sets the non-reciprocity $v+\gamma$; intercell bonds connect to neighbouring cells and remain purely passive.}
\label{fig:circuit}
\end{figure}
\section{\label{s5_expt}Proposed Experimental Platform: A Topolectrical Circuit}
In order to bridge our theoretical construction with a concrete experimental setting, we outline a topolectrical circuit (TEC) that emulates the driven BBH model \cite{Zhang2025, Di2026}. A TEC maps a tight-binding Hamiltonian onto the circuit Laplacian $J(\omega)$ via Kirchhoff's current law, such that at a chosen resonance frequency $\omega_0$ the admittance eigenvalues $j_n$ and eigenmodes reproduce the Hamiltonian spectrum and wavefunctions. Fixing the intercell hopping $\lambda$ to a reference capacitance $C_\lambda$, every other bond amplitude $t$ is realized by $C_t = (t/\lambda)C_\lambda$, so that the element-by-element dictionary reads as follows:
\begin{itemize}[leftmargin=1.5em]
  \item intercell $+\lambda$: capacitor $C_\lambda$, admittance $i\omega_0 C_\lambda$;
  \item intercell $-\lambda$ ($\pi$-flux bond): inductor $L_\lambda = 1/(\omega_0^2 C_\lambda)$, admittance $-i\omega_0 C_\lambda$;
  \item intracell $+v$: capacitor $C_v = (v/\lambda)C_\lambda$, admittance $i\omega_0 C_v$;
  \item intracell $-v$ ($\pi$-flux bond): inductor $L_v = 1/(\omega_0^2 C_v)$, admittance $-i\omega_0 C_v$;
  \item non-reciprocity $\pm\gamma$: an INIC $C_\gamma = (\gamma/\lambda)C_\lambda$ placed in parallel with each intracell bond, presenting admittance $i\omega_0(C_v \mp C_\gamma)$ in the two directions.
\end{itemize}
In short, a positive hopping $+t$ is realized by a capacitor with admittance $Y=i\omega_0 C_t$, while a negative hopping $-t$ (required for the $\pi$-flux plaquettes of the BBH lattice) is realized by an inductor $L_t = 1/(\omega_0^2 C_t)$, whose admittance $-i\omega_0 C_t$ is exactly its negative; the INIC is the element that renders the Laplacian non-Hermitian and physically generates the skin effect. Fig.~\ref{fig:circuit} shows the resulting unit cell.

The driven model, $U(T) = e^{-iH_2T/2}e^{-iH_1T/2}$, alternates an intracell non-Hermitian step $H_1$ with an intercell step $H_2$, and can be realized by time-multiplexing the same circuit: an analog switch (e.g., a CMOS transmission gate) in series with every bond, driven by a two-phase clock of period $T$, closes only the intracell elements ($C_v,L_v$) during the first half-period and only the intercell elements ($C_\lambda,L_\lambda$) and the intracell INICs during the second. Over one clock cycle the node voltages then evolve as $V \to U(T)V$, so the circuit stroboscopically implements the Floquet operator, with the clock frequency chosen well below $\omega_0$ so that each half-period remains adiabatic on the carrier time scale. Sampling the voltage profile at $t=NT$ images the Floquet skin modes directly, allowing this switched network to reveal the unipolar--bipolar crossover central to our proposal.

\section{\label{s6}Conclusion}

In this work, we have shown how periodic driving reshapes the symmetry structure and topological properties of a higher-order lattice system. Starting from the BBH model with an embedded $\pi$-flux configuration, we demonstrated that the static lattice realizes a projective form of space-time inversion symmetry, where the effective $\mathcal{PT}$ algebra mimics that of a spinful system despite the absence of physical spin degrees of freedom. Nevertheless, this symmetry transmutation alone does not activate first-order topology in the static limit, leaving the system confined to a purely higher-order phase with corner-localized states.
\par We showed that periodic driving provides a mechanism to overcome this constraint. By implementing a step-driving protocol that preserves global $\mathcal{PT}$ symmetry while dynamically modifying its algebraic structure, the driven lattice undergoes an interconversion of $\mathcal{PT}$-symmetric topological classes. As a result, the Floquet system realizes a hybrid-order topological phase where dispersive edge states coexist with symmetry-protected corner modes occupying distinct quasienergy sectors. Moreover, we have employed the spectral localizer as a real-space topological marker to identify these edge modes and have demonstrated their robustness against genuine random disorder.
\par Further, extending the analysis to the NH regime, we further uncovered a rich landscape of higher-order skin phenomena controlled by periodic driving. The interplay between non-reciprocity, crystalline symmetry, and the drive-induced symmetry modification produces transitions between unipolar and bipolar localization regimes and even realizes an anomalous $\mathbb{Z}_2$-type skin effect within an effectively spinless lattice. This crossover from a unipolar to a $\mathbb{Z}_2$-like bipolar phase enables a controllable delocalization--localization mechanism through drive-induced skin-suppressed points. Additionally, we have demonstrated that the bipolar/$\mathbb{Z}_2$-like localization is not merely a numerical artifact, but rather a symmetry-protected phenomenon that can be characterized by a symmetry-based topological index, namely the trace winding index.
\par In order to properly characterize these phenomena, we developed a non-Bloch Floquet framework based on a GBZ construction. By exploiting the underlying mirror symmetry of the model and restricting the analysis to the corresponding high-symmetry line, the two-dimensional problem can be effectively reduced to a one-dimensional representation. This enables the definition of a mirror-graded winding number and the construction of a 1D-mapped version of the 2D GBZ, from which a pair of non-Bloch invariants is obtained. These invariants correctly capture the emergence of corner states at quasienergies zero and $\pi/T$, thereby restoring the BBC in the periodically driven NH system.
\par Overall, our results demonstrate that Floquet engineering provides a powerful route for dynamically reshaping symmetry classification, generating hybrid higher-order topology, and controlling NH spectral localization in driven quantum systems.
\section{Data Availability}
The data that supports the findings of this article cannot be made publicly available.
The data are available upon reasonable request from the authors.\\
\section{Acknowledgments}
KR and LK sincerely acknowledge Srijata Lahiri for fruitful discussions.  Also, KR acknowledges the research fellowship from the MoE, Government of India. This work is supported in part by The Scientific and Technological Research 
Council of T{\"u}rkiye (T{\"U}B{\.I}TAK) under Grant No. 125F435 and Turkish Academy of Sciences (TUBA) under Grant
No. AD-2026. SB acknowledges support from T{\"U}B{\.I}TAK-B{\.I}DEB program.

\onecolumngrid
\appendix
\section{\label{app1}Drive-induced Symmetry Transmutation}

In general, a driven effective Hamiltonian does not necessarily preserve all the symmetries intrinsic to the static case. To restore the required symmetry operations, we introduce the following unitary transformations,
\begin{equation}
F_1 = e^{-i\mathcal{H}_1 T_1/2},
\qquad
F_2 = e^{\,i\mathcal{H}_2 T_2/2},
\label{eq:symmetric_frame}
\end{equation}
which define a pair of symmetric time frames. Under these transformations, the Floquet operators can be written as, $U_j = F_j U(T) F_j^{-1}$ ($j=1,2$), where $U_1$ and $U_2$ are defined as,
\begin{align}
U_1 &= e^{-i\mathcal{H}_1 T_1/2} \, e^{-i\mathcal{H}_2 T_2} \, e^{-i\mathcal{H}_1 T_1/2}, \\
U_2 &= e^{-i\mathcal{H}_2 T_2/2} \, e^{-i\mathcal{H}_1 T_1} \, e^{-i\mathcal{H}_2 T_2/2}.
\label{eq:app1_1-2}
\end{align}
Now, the effective Hamiltonian obtained in each of these frames satisfies the symmetry relations explicitly. For the time-reversal symmetry $\mathcal{T}$, we obtain
\begin{align}
\mathcal{T} U_1 \mathcal{T}^{-1} &= e^{i \mathcal{H}_1(-\mathbf{k}) T_1/2} \, e^{i \mathcal{H}_2(-\mathbf{k}) T_2} \, e^{i \mathcal{H}_1(-\mathbf{k}) T_1/2} = U_1^{\dagger}(-k), \\
\mathcal{T} U_2 \mathcal{T}^{-1} &= e^{i \mathcal{H}_2(-\mathbf{k}) T_2/2} \, e^{i \mathcal{H}_1(-\mathbf{k}) T_1} \, e^{i \mathcal{H}_2(-\mathbf{k}) T_2/2} = U_2^{\dagger}(-k),
\end{align}
which implies that
\begin{equation}
\mathcal{T} \mathcal{H}_{\mathrm{eff},j} \mathcal{T}^{-1} = \mathcal{H}_{\text{eff}}^{\dagger}(-\mathbf{k}), \quad (j=1,2).
\end{equation}
Similarly, for the chiral symmetry $S$, we find
\begin{align}
S U_1 S^{-1} &= e^{i \mathcal{H}_1(\mathbf{k}) T_1/2} \, e^{i \mathcal{H}_2(\mathbf{k}) T_2} \, e^{i \mathcal{H}_1(\mathbf{k}) T_1/2} = U_1^{-1}, \\
S U_2 S^{-1} &= e^{i \mathcal{H}_2(k) T_2/2} \, e^{i \mathcal{H}_1(\mathbf{k}) T_1} \, e^{i \mathcal{H}_2(\mathbf{k}) T_2/2} = U_2^{-1},
\end{align}
which leads to
\begin{equation}
S \mathcal{H}_{\mathrm{eff},j} S^{-1} = -\mathcal{H}_{\mathrm{eff},j}(\mathbf{k}).
\end{equation}
Thus, while the effective Hamiltonians in the symmetric time frames recover the required symmetries, the original effective Hamiltonian $\mathcal{H}_{\mathrm{eff}}$ does not explicitly exhibit them. Nevertheless, one can recover the hidden symmetry structures by defining the Floquet-renormalized symmetry operators as
\begin{equation}
\mathcal{T}_F = F_j^{-1}(-\mathbf{k})\,\mathcal{T}\,F_j(\mathbf{k}),
\qquad
S_F = F_j^{-1}(\mathbf{k})\,S\,F_j(\mathbf{k}),
\end{equation}
under which the effective Hamiltonian satisfies
\begin{align}
\mathcal{T}_F \mathcal{H}_{\mathrm{eff}}(\mathbf{k}) \mathcal{T}_F^{-1}
&= \mathcal{H}_{\mathrm{eff}}^{\dagger}(-\mathbf{k}), \\
S_F \mathcal{H}_{\mathrm{eff}}(\mathbf{k}) S_F^{-1}
&= -\mathcal{H}_{\mathrm{eff}}(\mathbf{k}).
\end{align}
This shows that the periodically driven system retains the essential symmetry features of its static counterpart, albeit in a dynamically renormalized form. However, the Floquet-renormalized symmetry operators no longer satisfy the commutation relations required by the underlying $\mathbb{Z}_2$ gauge structure (see Eq.~\ref{eq:gauge_condition}). In particular, $[G,\mathcal{T}_F] \neq 0$, implying that the projective inversion symmetry $\tilde{P} = G\mathcal{P}$ is no longer a valid symmetry in the driven system. As a consequence, the periodic drive modifies the $\mathcal{PT}$ algebra through the breakdown of the $\mathbb{Z}_2$ gauge structure, leading to the emergence of helical edge modes in a regime where the static model supports only higher-order topology.
\section{\label{app2}Effective $\mathbf{\textit{d}}$-vectors and Mirror-graded Winding Number}
As discussed in the main text, the static Hamiltonian preserves the mirror symmetry, which allows it to be expressed in a block-diagonal form along the symmetry line $k_x = k_y = k$. In particular, one can write
\begin{equation}
Q^{-1} H(k,k) Q =
\begin{pmatrix}
H^{+}(k) & 0 \\
0 & H^{-}(k)
\end{pmatrix},
\end{equation}
where $H^{+}(k)$ and $H^{-}(k)$ act on the mirror subspaces with eigenvalues $+1$ and $-1$, respectively. The corresponding unitary transformation is given by
\begin{equation}
Q =
\begin{pmatrix}
0 & 0 & 1 & 0 \\
1 & 0 & 0 & 0 \\
0 & \tfrac{1}{\sqrt{2}} & 0 & \tfrac{1}{\sqrt{2}} \\
0 & \tfrac{1}{\sqrt{2}} & 0 & -\tfrac{1}{\sqrt{2}}
\end{pmatrix}.
\end{equation}
The explicit form of each block can be written as
\begin{equation}
H^{\alpha}(k)
= \sqrt{2}\,\bigl[v + \lambda \cos k \bigr] \, \sigma_x
+ \alpha \sqrt{2}\,\bigl[\lambda \sin k + i \gamma \bigr] \, \sigma_y,
\qquad (\alpha = \pm),
\end{equation}
which can be recast into a massless Dirac form,
\begin{equation}
H^{\alpha}(k) = \mathbf{d}^{\alpha}(k) \cdot \boldsymbol{\sigma},
\end{equation}
with
\begin{equation}
\mathbf{d}^{\alpha}(k) = \bigl(d_x(k),\, \alpha\,  d_y(k), 0\bigr),
\end{equation}
where
\begin{equation}
d_x(k) = \sqrt{2}\,\bigl[v + \lambda \cos k \bigr], 
\qquad
d_y(k) = \sqrt{2}\,\bigl[\lambda \sin k + i \gamma \bigr].
\end{equation}
Moreover, such a Dirac Hamiltonian can be characterized by a winding number defined as,
\begin{equation}
\nu^{\alpha} = \frac{1}{2\pi i} \int_{0}^{2\pi}
\frac{1}{\mathbf{L}^{\alpha}(k)} 
\frac{d \mathbf{L}^{\alpha}(k)}{dk} \, dk,
\label{eq:winding}
\end{equation}
with
\begin{equation}
\mathbf{L}^{\alpha}(k) = d_x(k) + \alpha\,i\,  d_y(k).
\end{equation}
The higher-order topology of the system can now be captured by the mirror-graded invariant
\begin{equation}
\nu = \frac{\nu^{+} - \nu^{-}}{2}.
\end{equation}
Importantly, the presence of the NH parameter $\gamma$ induces the skin effect, which breaks the conventional BBC. To restore the BBC, one must employ non-Bloch band formalism, where the topology is defined over the GBZ. This can be implemented by replacing $e^{ik}$ with $\beta$, and correspondingly deforming the integration contour from $[0,2\pi]$ in Eq. \ref{eq:winding} to the GBZ trajectory $\mathcal{C}_\beta$. A convenient approximation is to introduce a complex momentum $k \rightarrow \tilde{k} = k - i \ln r$, where \cite{Wang2021},
\begin{equation}
r = |\beta| = \left|\frac{v-\gamma}{v+\gamma}\right|^{1/2}.
\end{equation}
In the periodically driven case, however, this procedure cannot be applied directly due to the loss of symmetry in the effective Floquet Hamiltonian. As discussed in Appendix. \ref{app1}, one must instead work within the symmetric time frames $U_1$ and $U_2$. 
Within each frame $(j=1,2)$, the effective Hamiltonian can be extracted from the Floquet evolution operator in the following form,
\begin{equation}
    U_{j}^{\alpha} = \text{exp}[-i E(\beta) \mathbf{n}_j^{\alpha}(\beta) \cdot \sigma],
\end{equation}
where both the $U_{j}^{\alpha}$ are expressed along the symmetry line $k_x=k_y=k$.
Exploiting the closed algebra and the Euler decomposition of Pauli matrices, the Floquet evolution operator in each symmetric frame can be written as
\begin{equation} U_j(k) \;=\; \cos E\;\mathbbm{1} \;-\; i\,\sin E\;\mathbf{n}_j\!\cdot\!\boldsymbol{\sigma}. \label{eq:product} \end{equation}
with $\mathbf{n}_j$ being the corresponding effective Bloch vector in the $j$th frame, and can be expressed as,
\begin{align}
\mathbf{n}_j &= 
\Bigl[
\cos(|\mathcal{H}_j|T_j)\sin(|\mathcal{H}_i|T_i)\,
\hat{\mathcal{H}}_j \cdot \hat{\mathcal{H}}_i
+ \sin(|\mathcal{H}_j|T_j)\cos(|\mathcal{H}_i|T_i)
\Bigr] \hat{\mathcal{H}}_j \nonumber \\
&\quad + \sin(|\mathcal{H}_i|T_i)
\Bigl[
\hat{\mathcal{H}}_i
- (\hat{\mathcal{H}}_j \cdot \hat{\mathcal{H}}_i)\hat{\mathcal{H}}_j
\Bigr],
\end{align}
with $(i,j) = (1,2)$ or $(2,1)$ labelling the two symmetric frames. It is this vector, obtained
directly from the product of the step propagators, that we identify as the effective $d$-vector in
frame $j$. Substituting the explicit forms of $\mathcal{H}_1^{\alpha}(k)$ and $\mathcal{H}_2^{\alpha}(k)$, one obtains the corresponding Bloch vectors $\mathbf{n}_j^{\alpha}$ in each frame, which can be used to evaluate the winding numbers for each mirror sector $\alpha$ in each symmetric frame $j$. The integration must again be performed over the Floquet GBZ, whose construction has been outlined in the main text. Finally, a complete characterization of the Floquet higher-order topology requires two independent invariants, $\nu_{0}$ and $\nu_{\pi}$, corresponding to corner modes at $0$ and $\pi/T$ quasienergies, respectively. These are obtained by combining the winding numbers $\nu_{1}$ and $\nu_{2}$ (evaluated in the two symmetric frames) according to the relation given in Eq.~\ref{eq:winding_two_frames} of the main text.
\twocolumngrid

\bibliography{ref}
\end{document}